\documentclass[aps,prd,preprint,showpacs,eqsecnum,nofootinbib]{revtex4}
\usepackage{graphicx,epsf,bm}
\newif\ifdraft
\drafttrue
\newif\ifpreprint
\preprinttrue

\def\sect#1{section~{\ref{#1}}}

\def\spa#1.#2{\left\langle#1\,#2\right\rangle}
\def\spb#1.#2{\left[#1\,#2\right]}
\def\spash#1.#2{\spa{\smash{#1}}.{\smash{#2}}}
\def\spbsh#1.#2{\spb{\smash{#1}}.{\smash{#2}}}
\def\lor#1.#2{\left(#1\,#2\right)}
\def\sand#1.#2.#3{%
\left\langle\smash{#1^{-}}{\vphantom1}\right|{#2}%
\left|\smash{#3^{-}}{\vphantom1}\right\rangle}
\def\sandpp#1.#2.#3{%
\left\langle\smash{#1^{+}}{\vphantom1}\right|{#2}%
\left|\smash{#3^{+}}{\vphantom1}\right\rangle}
\def\sandpm#1.#2.#3{%
\left\langle\smash{#1^{+}}\vphantom1\right|{#2}%
\left|\smash{#3^{-}}\vphantom1\right\rangle}
\def\sandmp#1.#2.#3{%
\left\langle\smash{#1^{-}}\vphantom1\right|{#2}%
\left|\smash{#3^{+}}{\vphantom1}\right\rangle}
\def\sandmppm#1.#2.#3{%
\left\langle\smash{#1^{\mp}}\vphantom1\right|{#2}%
\l
eft|\smash{#3^{\pm}}{\vphantom1}\right\rangle}
\def\sandnn#1.#2.#3{%
\left\langle\smash{#1}\vphantom1\right|{#2}%
\left|\smash{#3}{\vphantom1}\right\rangle}
\def\sandmn#1.#2.#3{%
\left\langle\smash{#1^{-}}\vphantom1\right|{#2}%
\left|\smash{#3}{\vphantom1}\right\rangle}
\def\sandnm#1.#2.#3{%
\left\langle\smash{#1}\vphantom1\right|{#2}%
\left|\smash{#3^{-}}{\vphantom1}\right\rangle}

\def\e{\epsilon}
\def\eps{\epsilon}

\def\nn{\nonumber}

\def\eqn#1{eq.~(\ref{#1})}

\def\NeqFour{{{\cal N}=4}}

\def\OneLoop{I}
\def\OneLoopProduct{I}

\def\be{\begin{equation}}
\def\ee{\end{equation}}
\def\bea{\begin{eqnarray}}
\def\eea{\end{eqnarray}}
\def\ba{\begin{eqnarray}}
\def\ea{\end{eqnarray}}
\def\bm{\boldmath}

\def\tcdot{\!\cdot\!}

\def\Ord{{\cal O}}

\newbox\charbox
\newbox\slabox
\def\s#1{{      
        \setbox\charbox=\hbox{$#1$}
        \setbox\slabox=\hbox{$/$}
        \dimen\charbox=\ht\slabox
        \advance\dimen\charbox by -\dp\slabox
        \advance\dimen\charbox by -\ht\charbox
        \advance\dimen\charbox by \dp\charbox
        \divide\dimen\charbox by 2
        \raise-\dimen\charbox\hbox to \wd\charbox{\hss/\hss}
        \llap{$#1$} }}

\begin{document}

\ifpreprint
Saclay-IPhT--T10/089
\qquad
WIS/09/10-JULY-DPPA
\fi

\title{Towards a Basis for Planar Two-Loop Integrals}

\author{Janusz Gluza}\affiliation{
Department of Field Theory and Particle Physics, Institute of Physics,
University of Silesia, Uniwersytecka 4, PL--40-007 Katowice, Poland
}
\author{Krzysztof Kajda}
\affiliation{
Department of Field Theory and Particle Physics, Institute of Physics,
University of Silesia, Uniwersytecka 4, PL--40-007 Katowice, Poland
}\author{David A. Kosower}\affiliation{Institut de Physique Th\'eorique, CEA--Saclay,
          F--91191 Gif-sur-Yvette cedex, France}
\affiliation{Department of Particle Physics and Astrophysics, Weizmann Institute of Science, Rehovot 76100, Israel}

\begin{abstract}
The existence of a finite basis of algebraically independent 
one-loop integrals has underpinned important developments in
the computation of one-loop amplitudes in field theories and gauge
theories in particular.  We give an explicit construction
reducing integrals with massless propagators to a finite basis for 
planar
integrals at two loops, both to all orders in the dimensional
regulator $\e$, and also when all integrals are truncated to
$\Ord(\e)$.  We show how to reorganize integration-by-parts
equations to obtain elements of the first basis efficiently,
and how to use Gram determinants
to obtain additional linear relations reducing this
all-orders basis to the second one.  The techniques we present
should apply to non-planar integrals, to integrals with massive
propagators, 
and beyond two loops as well.
\end{abstract}


\maketitle

\section{Introduction}
\label{IntroSection}

\def\AIR{{\sc air}}
\def\FIRE{{\sc Fire}}
\def\Reduze{{\tt Reduze}}
The computation of higher-order corrections to amplitudes in gauge
theories is important to the experimental program at particle
colliders. Recent years have witnessed dramatic advances in 
technologies for computing one-loop amplitudes, critical to the
program of next-to-leading order calculations for collider physics.

  Important advances have also been made in computations
of amplitudes beyond one loop.
 The computation of two-loop amplitudes relies in part on
the ability to compute two-loop integrals, which has seen remarkable
progress in the last decade.  
Several technologies~\cite{SmirnovBook} played a role in these
advances, most notably the Mellin--Barnes approach
 to computing
integrals pioneered by Smirnov~\cite{SmirnovDoubleBox} and
Tausk~\cite{Tausk}, and later automated by Czakon~\cite{CzakonMB}.  
Smirnov and Smirnov have recently introduced an alternative 
automated strategy for resolving singularities~\cite{SmirnovStrategy}.
Anastasiou et al.{} have developed another
method~\cite{ContourDeformationCombined} of integral evaluation combining
sector decomposition~\cite{SectorDecomposition} with 
contour deformation~\cite{ContourDeformation}.
These 
technologies have played a key role in higher-loop
calculations in the $\NeqFour$ supersymmetric 
gauge theory~\cite{Neq4Calculations,LeadingSingularityCalculation}.  

The computation of amplitudes has also made use of techniques for
reducing arbitrary tensor integrals to a basis set of scalar
{\it master
  integrals\/}.  In calculations performed to date,
the reductions have
relied on integration by parts (IBP)~\cite{IBP} to construct
linear equations relating the various integrals, and on Gaussian elimination
in the form of the Laporta algorithm~\cite{Laporta}
 to solve them.  The solution
determines a set of master integrals, and gives expressions for the
remaining integrals in terms of them.  This reduction approach
has been automated in Anastasiou and Lazopoulos's \AIR{}
program~\cite{AIR}, in Smirnov's \FIRE{} program~\cite{FIRE},
and more recently, in Studerus's \Reduze{} program~\cite{Reduze},
as well as various private computer codes.
We should note that the existence of
a method, such as the Mellin--Barnes approach, for evaluating loop integrals
directly  means that a reduction to master integrals
is not, strictly speaking, necessary for a Feynman-diagram
calculation.  It greatly reduces the
complexity and difficulty of such calculations, however.
In order to use master integrals in such calculations, one needs the explicit
forms of the reduction equations.

Recent years have also witnessed the development and elaboration of a
new set of technologies, so-called on-shell
methods~\cite{UnitarityMethod,BCFUnitarity,OnShellMethods,Forde,%
BCFW,Bootstrap,DdimUnitarity}, 
for computing amplitudes.  These rely
only on knowledge gleaned from physical states.  The unitarity method,
one of the tools in this approach, determines the rational
coefficients of loop integrals in terms of products of tree amplitudes
corresponding to cutting propagators in the loop amplitude.
(These coefficients are rational in spinor variables.)

It is possible to determine the set of loop integrals that contribute
to a given process during the computation of their coefficients,
and most of the higher-loop computations to date have proceeded in this
manner.  In the most powerful form of the unitarity method, 
generalized unitarity~\cite{Zqqgg,BCFUnitarity,TwoLoopSplitting,FiveLoop}, one
cuts an amplitude into more than two pieces; indeed, in
`maximal unitarity,' one cuts as many
propagators as possible in a given contribution, thereby reducing 
any higher-loop amplitude to a product of basic tree amplitudes.  The power of
this technique is greatly enhanced by an {\it a priori\/} knowledge
of a basis of integrals, as it then becomes possible to design the cuts in a
general way.  

Knowledge of a basis, in contrast, is essential to developing
an automated numerical implementation, which several groups
are currently pursuing at one loop~\cite{NumericalOnShell}.
The required basis has been known for a long time at one loop.
The four-dimensional one dates back to the work of 
Melrose~\cite{Melrose}.  
It is worth noting that the reduction equations
themselves are not required when using generalized unitarity,
because the method avoids the need for reductions of integrals with non-trivial numerators.  We only need to know the set of algebraically 
independent master integrals.  Baikov's work~\cite{Baikov} suggests
an interesting connection between integration by parts and 
maximal unitarity.

Smirnov and Petukhov~\cite{SmirnovPetukhov}
have recently shown that the integral basis
resulting from integration by parts is finite.  
In this paper, we give an explicit reduction to a finite
set of integrals for planar
integrals at two loops.  There are two different kinds of bases we
will consider.  One requires algebraic independence to all orders in
the dimensional regulator $\e$ (a ``$D$-dimensional basis''), while the
other requires algebraic independence for integrals truncated to
$\Ord(\e^0)$ (a ``regulated four-dimensional basis'').  The latter
contains fewer integrals, and is the relevant basis for the
computation of amplitudes for numerical applications.  We shall show
how to limit the set of planar integrals that enter into a general
two-loop computation, and discuss the reductions of some of these
integrals.  We leave the complete enumeration of basis integrals, as
well as proofs of their algebraic independence, to future work.

The approach we will pursue here makes use of a chosen subset of IBP
equations, designed to avoid the introduction of unwanted integrals with
doubled propagators,
as well as supplementary Gram-determinant equations to take advantage
of additional reductions possible when the loop integrals are
performed in a truncated expansion about four dimensions rather than
in arbitrary dimension (that is, to all orders in the $\e$ expansion).
The approach we describe should also apply to non-planar integrals,
and beyond two loops as well.  We use the Mellin--Barnes 
approach~\cite{CzakonMB,SmirnovStrategy,Gluza:2007rt,Gluza:2010mz} to
cross-check our equations, along with another technique for evaluating
general higher-loop integrals, sector
decomposition~\cite{SectorDecomposition,Gluza:2010mz}.

We will not discuss the analytic evaluation of the master integrals.  Integrals
involving a single dimensionless ratio of invariants may be expressed
in terms of harmonic polylogarithms introduced by Vermaseren and
Remiddi~\cite{HarmonicPolylogarithms}, or alternatively in terms of
the generalized polylogarithms of Goncharov~\cite{Goncharov}; some
integrals involving two dimensionless ratios can be expressed in terms
of a two-dimensional generalization of harmonic polylogarithms
introduced by Gehrmann and Remiddi~\cite{GehrmannRemiddi}; for
examples, see ref.~\cite{TwoRatioExamples}.  The four-mass
double box was computed 
by Ussyukina and Davydychev~\cite{FourMassDoubleBox,DualConformalI}.
It is plausible
that the complete set of two-loop basis integrals with massless
internal lines can be
expressed in terms of generalized polylogarithms, but this remains
to be proven.

In section~\ref{OneLoopSection}, we review the basis of one-loop
integrals with massless propagators
in order to illustrate the two different bases, and
to give a simple example of the use
of Gram-determinant equations.
In section~\ref{PlanarSection}, we show how to reduce two-loop
tensor and scalar integrals of sufficiently high multiplicity
(again with massless propagators),
thereby providing a constructive demonstration
of the existence of a finite basis.  We also
describe how to obtain a compact set of equations relating 
only integrals relevant to amplitudes, avoiding the introduction
of integrals with doubled propagators.
In section~\ref{MasslessDoubleBoxSection}, we discuss the massless
double box in detail.
In section~\ref{MassiveDoubleBoxSection}, we apply these techniques
to double-box integrals with different patterns of external
masses.  In section~\ref{PentaboxSection}, we apply the
techniques to the pentabox integral.  
In section~\ref{SixPointSection},
we give one example of the reduction of a six-point integral, the
double pentagon.
In section~\ref{GeneralizedUnitaritySection}, we present
a heuristic explanation of some of our results using
generalized unitarity.  We summarize in a concluding
section.


\section{Reduction of One-Loop Integrals}
\label{OneLoopSection}

\begin{figure}
\begin{minipage}[b]{0.24\linewidth}
\centering
\includegraphics[scale=0.5]{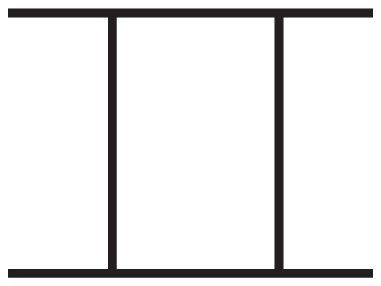}
\center{(a)}
\end{minipage}
\begin{minipage}[b]{0.24\linewidth}
\centering
\includegraphics[scale=0.5]{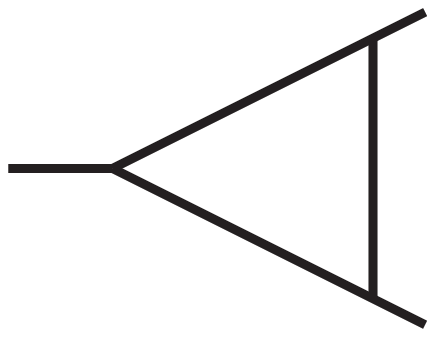}
\center{(b)}
\end{minipage}
\begin{minipage}[b]{0.24\linewidth}
\centering
\includegraphics[scale=0.5]{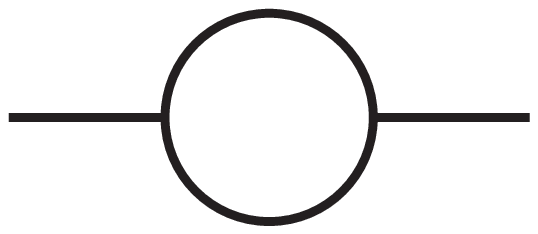}
\center{(c)}
\end{minipage}
\begin{minipage}[b]{0.24\linewidth}
\centering
\includegraphics[scale=0.5]{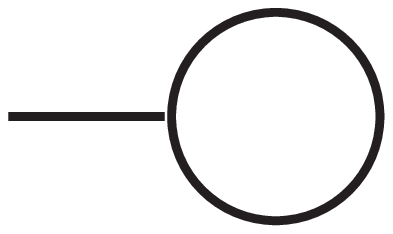}
\center{(d)}
\end{minipage}

\begin{center}
 \caption{The basis of scalar integrals: (a) box, (b) triangle, 
(c) bubble, and (d) tadpole.  Each corner can have one or
more external momenta emerging from it.  The tadpole integral (d) vanishes
when all internal propagators are massless.} 
\label{OneLoopIntegralBasisFigure}
\end{center}
\end{figure}

As a warm-up exercise, let us review integral bases at one loop along with their derivation.  Throughout the paper, we will take the
external momenta to be strictly four-dimensional.  They may be
massless, or massive (representing, for example, sums of massless
momenta in the original amplitude).  In addition, we will take all
vectors contracted with the loop momentum to be strictly
four-dimensional as well.  These vectors might be momenta or
polarization vectors.  All internal lines are taken to be
massless.

In an $n$-point one-loop amplitude in gauge theory, we start with
integrals with up to $n$ external legs, and up to $n$ powers of the
loop momentum in the numerator.  (In a gravitational theory, we would
start with up to $2n$ powers of the loop momentum.  Up to questions
regarding ultraviolet divergences, their treatment follows the same
approach as the gauge-theory tensor integrals.)  These powers are
contracted with external momenta, external polarization vectors, or
external currents.  We shall denote the scalar integral
 by $\OneLoop_n$,
\begin{equation}
\OneLoop_n(K_1,\ldots,K_n) \equiv \OneLoop_n[1] \equiv -i\int {d^D\ell\over(2\pi)^D}\; 
   {1\over \ell^2 (\ell-K_1)^2 (\ell-K_{12})^2\cdots (\ell-K_{1\cdots(n-1)})^2}\,.
\end{equation}
 In this equation, $K_{j\cdots l} =
K_{j}+\cdots+K_l$.  We denote integrals with a function of $\ell$
inserted in the numerator as follows,
\begin{equation}
\OneLoop_n[{\cal P}(\ell)] \equiv -i\int {d^D\ell\over(2\pi)^D}\; 
   {{\cal P}(\ell)\over \ell^2 (\ell-K_1)^2 (\ell-K_{12})^2\cdots (\ell-K_{1\cdots(n-1)})^2}\,,
\end{equation}
where the momentum arguments are left implicit.

\def\GramO#1{G(#1)} \def\Gram#1#2{G\biggl({#1\atop#2}\biggr)} 
\def\GramSq#1#2{G^2\biggl({#1\atop#2}\biggr)} 

Let us begin with high-multiplicity integrals, with five or more
external legs.  Consider a generic tensor integral\footnote{While this
numerator does not have free indices, they could be exhibited by
differentiating with respect to the $v_j$, so in a slight
abuse of language, we will refer to the
integral as a tensor integral.}  $\OneLoop_n[\ell\cdot
v_1\,\ell\cdot v_2\cdots\ell\cdot v_n]$.  We shall make
use of Gram determinants,
\begin{eqnarray}
\Gram{p_1,\cdots,p_l}{q_1,\cdots,q_l} &\equiv& \det_{i,j\in l\times l}(2 p_i\cdot q_j)\,,\\
\GramO{p_1,\cdots,p_l} &\equiv& \Gram{p_1,\cdots,p_l}{p_1,\cdots,p_l}\,,
\end{eqnarray}
which have the useful property that they vanish if either the $\{p_j\}$ or
the $\{q_j\}$ are linearly dependent.  Using these objects,
 we can expand each of
the four-dimensional vectors $v_j$ in a basis of four chosen external momenta $b_1, b_2, b_3, b_4$,
\begin{eqnarray}
v^\mu &=& \frac{1}{\GramO{b_1,b_2,b_3,b_4}}\biggl[
 \Gram{v,b_2,b_3,b_4}{b_1,b_2,b_3,b_4} b_1^\mu
 +\Gram{b_1,v,b_3,b_4}{b_1,b_2,b_3,b_4} b_2^\mu
 +\Gram{b_1,b_2,v,b_4}{b_1,b_2,b_3,b_4} b_3^\mu\nonumber\\
&&\hphantom{ \frac{1}{\GramO{b_1,b_2,b_3,b_4}}\biggl[}
    +\Gram{b_1,b_2,b_3,v}{b_1,b_2,b_3,b_4} b_4^\mu
\biggr]\,.
\label{MomentumExpansion}
\end{eqnarray}
This leads us to consider integrals with numerator insertions where
all of the $v_j$ are equal to one of the $b_i$.  At one loop,
these factors are all
reducible, because we can rewrite any dot product as a difference of
denominators, for example,
\begin{equation}
\ell\cdot b_1 = \frac12 \bigl[ (\ell-K)^2 - (\ell-K-b_1)^2 + (K+b_1)^2 - K^2\bigr]\,.
\label{ReducibleNumerator}
\end{equation}
The first two terms lead to integrals with fewer propagators and fewer
powers of $\ell$ in the numerator, while the last two lead to
integrals with fewer powers of $\ell$,
\begin{equation}
\OneLoop_n[(\ell\cdot v)^n] \longrightarrow \OneLoop_{n-1}[(\ell\cdot v)^{n-1}] \oplus \OneLoop_n[(\ell\cdot v)^{n-1}]\,.
\end{equation}
Repeating this procedure for the daughter integrals (with a new basis
element instead of $b_1$ where required) ultimately leads to integrals
$\OneLoop_n$ with $n\leq 4$ or with trivial numerators.  

\def\tb{\tilde b}
As is well known~\cite{IntegralReductionsI,Melrose,IntegralReductionsII,IntegralReductionsIII}, 
we can also reduce four- or fewer-point integrals
with non-trivial numerators, by relying on Lorentz invariance
to re-express them in terms of integrals where the non-trivial numerators
involve only external momenta.  (We could alternatively introduce additional
basis vectors up to contributions from $\Ord(\e)$ numerators.)
 For the purposes of
studying reductions, it therefore suffices to take $v$ to be
one of the external momenta even though the latter do not suffice to 
provide a basis.  At higher loops, not all integrals can be reduced
this way, because not all numerators can be written as differences
of propagator denominators as in \eqn{ReducibleNumerator}.
Many integrals with irreducible numerators can
nonetheless be simplified using IBP technology as implemented 
(for example) in the Anastasiou--Lazopoulos \AIR{} code~\cite{AIR}
in Smirnov's \FIRE{} package~\cite{FIRE}, 
or in Studerus's \Reduze{} package~\cite{Reduze},
leaving a smaller
set of master integrals.

We must next reduce the five- or higher-point integrals with trivial
numerators (`scalar' integrals).  While the reductions above hold
independent of the dimensionality of the loop integration, the same is
not true of all of the reductions we must consider here.  We must
distinguish between an integral basis to all orders in the
dimensional regularization parameter $\e$, and one which holds only
through order $\Ord(\e^0)$.  The latter may contain fewer integrals than
the former, because it is possible for linear combinations of
integrals to be nonzero but of $\Ord(\e)$.  At one loop, this is
indeed what happens, with the scalar pentagon integral required in an
all-orders basis while being reducible to $\Ord(\e^0)$~\cite{FivePoint}.

\def\nn{\nonumber} 
Let first consider the reduction of six- or
higher-point integrals, which can be done to all orders in $\e$.
Because the external momenta are four-dimensional, we have,
\begin{equation}
\Gram{\ell,1,2,3,4}{5,1,2,3,4} = 0\,,
\end{equation}
where we have used the labels of the external momenta to represent the momenta themselves.
Accordingly,
\begin{equation}
\OneLoop_n\Bigl[\Gram{\ell,1,2,3,4}{5,1,2,3,4}\Bigr] = 0\,,\hskip 1cm (n\geq 6)\,.
\end{equation}
If we now expand the Gram determinant,
\begin{eqnarray}
\Gram{\ell,1,2,3,4}{5,1,2,3,4} 
&=& 
-\ell^2\,\Gram{1,2,3,4}{5,2,3,4}
+(\ell-K_1)^2\,\Gram{1,2,3,4}{5,K_{12},3,4}
\nn\\&&
-(\ell-K_{12})^2\,\Gram{1,2,3,4}{5,1,K_{23},4}
+(\ell-K_{123})^2\,\Gram{1,2,3,4}{5,1,2,K_{34}}
\nn\\&&
+(\ell-K_{1234})^2\,\Gram{1,2,3,4}{1,2,3,K_{45}}
- (\ell-K_{12345})^2\, \Gram{1,2,3,4}{1,2,3,4}
\nn\\&&
-K_1^2\,\Gram{1,2,3,4}{5,K_{12},3,4}
  +K_{12}^2\,\Gram{1,2,3,4}{5,1,K_{23},4}
\nn\\&&
-K_{123}^2\,\Gram{1,2,3,4}{5,1,2,K_{34}}
  -K_{1234}^2\,\Gram{1,2,3,4}{1,2,3,K_{45}}
\nn\\&&
+K_{12345}^2\,\Gram{1,2,3,4}{1,2,3,4}\,,
\end{eqnarray}
we obtain an equation relating the $n$-point integral to 
six $(n-1)$-point integrals,
\begin{eqnarray}
\OneLoop_n(K_1,\ldots,K_n) &=& c_1 \OneLoop_{n-1}(K_{n1},K_2,\ldots,K_{n-1})
                   +c_2 \OneLoop_{n-1}(K_{12},K_3,\ldots,K_{n})\nn\\
                &&\hskip -15mm  +c_3 \OneLoop_{n-1}(K_{1},K_{23},K_4,\ldots,K_{n})
                      +c_4 \OneLoop_{n-1}(K_{1},K_2,K_{34},K_5,\ldots,K_{n})\\
                &&\hskip -15mm  +c_5 \OneLoop_{n-1}(K_{1},\ldots,K_{45},\ldots,K_{n})
                      +c_6 \OneLoop_{n-1}(K_{1},\ldots,K_{56},\ldots,K_{n})\,\nn
\end{eqnarray}
where
\begin{eqnarray}
c_0 &=&
-K_1^2\,\Gram{1,2,3,4}{5,K_{12},3,4}
  +K_{12}^2\,\Gram{1,2,3,4}{5,1,K_{23},4}
-K_{123}^2\,\Gram{1,2,3,4}{5,1,2,K_{34}}
\nn\\&&
-K_{1234}^2\,\Gram{1,2,3,4}{1,2,3,K_{45}}
+K_{12345}^2\,\Gram{1,2,3,4}{1,2,3,4}\,,\nn\\
c_1 &=& 
{1\over c_0}\,\Gram{1,2,3,4}{5,2,3,4}\,,\nn\\
c_2 &=& 
-{1\over c_0}\,\Gram{1,2,3,4}{5,K_{12},3,4}\,,
\label{HexagonReductionCoefficients}\\
c_3 &=& 
{1\over c_0}\,\Gram{1,2,3,4}{5,1,K_{23},4}\,,\nn\\
c_4 &=& 
-{1\over c_0}\,\Gram{1,2,3,4}{5,1,2,K_{34}}\,,\nn\\
c_5 &=& 
-{1\over c_0}\,\Gram{1,2,3,4}{1,2,3,K_{45}}\,,\nn\\
c_6 &=& 
{1\over c_0}\, \Gram{1,2,3,4}{1,2,3,4}\,.\nn
\end{eqnarray}
One can check numerically that 
the coefficient $c_0$ does not
vanish for generic momenta, and hence the $c_i$ are well-defined.

In $D$ dimensions, as mentioned above, pentagon integrals are needed
as independent basis elements.  When expanding about $D=4-2\e$
dimensions, however, only the $\Ord(\e)$ terms are independent, so
that the integral can be eliminated from the basis.  We can derive the
required equation by considering the Gram determinant
$\GramO{\ell,1,2,3,4}$.  The Gram determinant itself is of $\Ord(\e)$,
because it can avoid vanishing only when the $\e$ components of $\ell$
appear in place of $\ell$.  This leads us to consider the integral,
\begin{equation}
\OneLoop_5[\GramO{\ell,1,2,3,4}]\,.
\end{equation}
One might worry that the Gram determinant can end up multiplying
divergent terms in the integrand, yielding terms which are overall
still of $\Ord(\e^0)$ or even divergent.  However, all divergences of
the integral arise from regions where $\ell$ is soft, or collinear to
one of the external legs.  In these regions, the Gram determinant
vanishes.  Because the divergences are logarithmic at $D=4$, any vanishing of
the integrand suffices to eliminate the divergences.  (At one loop,
this in fact follows directly from the dependence of the integral only
on the $\e$-dimensional components of $\ell$; but that will not necessarily be
manifestly true for similar integrals we shall consider in the two-loop case.)
Furthermore, the integral is ultraviolet-finite by power counting.
Accordingly, the integral itself is also of $\Ord(\e)$,
\begin{equation}
\OneLoop_5[\GramO{\ell,1,2,3,4}] = \Ord(\e)\,.
\end{equation}
We can use this to obtain a useful equation for the pentagon integral by expanding
the Gram determinant, and re-expressing dot products of the loop momenta in terms
of differences of denominators,
\begin{eqnarray}
\GramO{\ell,1,2,3,4} 
&=&
d_0 + d_1 \ell^2 + d_2 (\ell-K_1)^2 + d_3 (\ell-K_{12})^2 + d_4 (\ell-K_{123})^2
+d_5 (\ell-K_{1234})^2
\nn\\
&&
-\ell^2\,\Gram{1,2,3,4}{\ell,2,3,4}
+(\ell-K_1)^2\,\Gram{1,2,3,4}{\ell,K_{12},3,4}
-(\ell-K_{12})^2\,\Gram{1,2,3,4}{\ell,1,K_{23},4}\nn\\
&&+(\ell-K_{123})^2\,\Gram{1,2,3,4}{\ell,1,2,K_{34}}
-(\ell-K_{1234})^2\,\Gram{1,2,3,4}{\ell,1,2,3}\,,
\label{Gram5a}
\end{eqnarray}
where
\begin{eqnarray}
d_0 &=& 
-(K_{1}^2)^{2} \Gram{K_{12}, 3, 4}{K_{12}, 3, 4}
        +2 K_{1}^2K_{12}^2 \Gram{K_{12}, 3, 4}{1,K_{23}, 4}
        -(K_{12}^2)^{2} \Gram{1, K_{23}, 4}{1, K_{23}, 4}
\nn\\ &&
        -2 K_{1}^2K_{123}^2 \Gram{K_{12}, 3, 4}{1, 2, K_{34}}
        +2 K_{12}^2K_{123}^2 \Gram{1, K_{23}, 4}{1, 2,K_{34}}
        -(K_{123}^2)^{2} \Gram{1, 2, K_{34}}{1, 2, K_{34}}
\nn\\ &&
        +2 K_{1}^2K_{1234}^2 \Gram{K_{12}, 3, 4}{1, 2, 3}
        -2 K_{12}^2K_{1234}^2 \Gram{1, K_{23}, 4}{1, 2,3}
        +2 K_{123}^2K_{1234}^2 \Gram{1, 2, K_{34}}{1, 2, 3}
\nn\\ &&
        -(K_{1234}^2)^{2} \Gram{1,2, 3}{1, 2, 3}
        \,,\nn\\
d_1 &=& 
2 \Gram{1,2,3,4}{1,2,3,4}
        - K_{1}^2 \Gram{K_{12},3,4}{2,3,4}
        +K_{12}^2 \Gram{1,K_{23},4}{2,3,4}
        - K_{123}^2 \Gram{1,2,K_{34}}{2,3,4}
\nn\\ &&
        +K_{1234}^2 \Gram{1,2,3}{2,3,4}\,,
\label{PentagonReductionCoeffs}\\
d_2 &=& 
K_{1}^2 \Gram{K_{12},3,4}{K_{12},3,4}
        -K_{12}^2 \Gram{K_{12},3,4}{1,K_{23},4}
        +K_{123}^2 \Gram{K_{12},3,4}{1,2,K_{34}}
        -K_{1234}^2 \Gram{K_{12},3,4}{1,2,3}\,,\nn\\
d_3 &=& 
-K_{1}^2 \Gram{K_{12},3,4}{1,K_{23},4}
        +K_{12}^2 \Gram{1,K_{23},4}{1,K_{23},4}
        -K_{123}^2 \Gram{1,K_{23},4}{1,2,K_{34}}
        +K_{1234}^2 \Gram{1,K_{23},4}{1,2,3}\,,\nn\\
d_4 &=& 
K_{1}^2 \Gram{K_{12},3,4}{1,2,K_{34}}
        -K_{12}^2 \Gram{1,K_{23},4}{1,2,K_{34}}
        +K_{123}^2 \Gram{1,2,K_{34}}{1,2,K_{34}}
        -K_{1234}^2 \Gram{1,2,K_{34}}{1,2,3}\,,\nn\\
d_5 &=& 
-K_{1}^2 \Gram{K_{12},3,4}{1,2,3}
        +K_{12}^2 \Gram{1,K_{23},4}{1,2,3}
        -K_{123}^2 \Gram{1,2,K_{34}}{1,2,3}
        +K_{1234}^2 \Gram{1,2,3}{1,2,3}\,.\nn
\end{eqnarray}
The integrals of the terms on the last two lines of \eqn{Gram5a} will 
vanish, as they correspond to box integrals with $\varepsilon(\ell,\ldots)$
in the numerator.  The integral of the $d_0$ term is simply a pentagon,
and the integrals of the $d_{1,\ldots,5}$ terms are box integrals.
These reductions and relations yield the well-known basis
shown in fig.~\ref{OneLoopIntegralBasisFigure}.

Our aim is to extend these considerations to two-loop integrals.
We will also introduce a new technique for reducing integrals with
irreducible numerators to the set of master
integrals, based on rewriting the system of IBP equations.
We delineate the finite universal basis,
while leaving a complete and detailed enumeration of it to future work.

\section{Reduction of Planar Two-Loop Integrals}
\label{PlanarSection}
\subsection{The Integrals}

\begin{figure}[ht]
\begin{minipage}[b]{0.4\linewidth}
\centering
\includegraphics[scale=0.5]{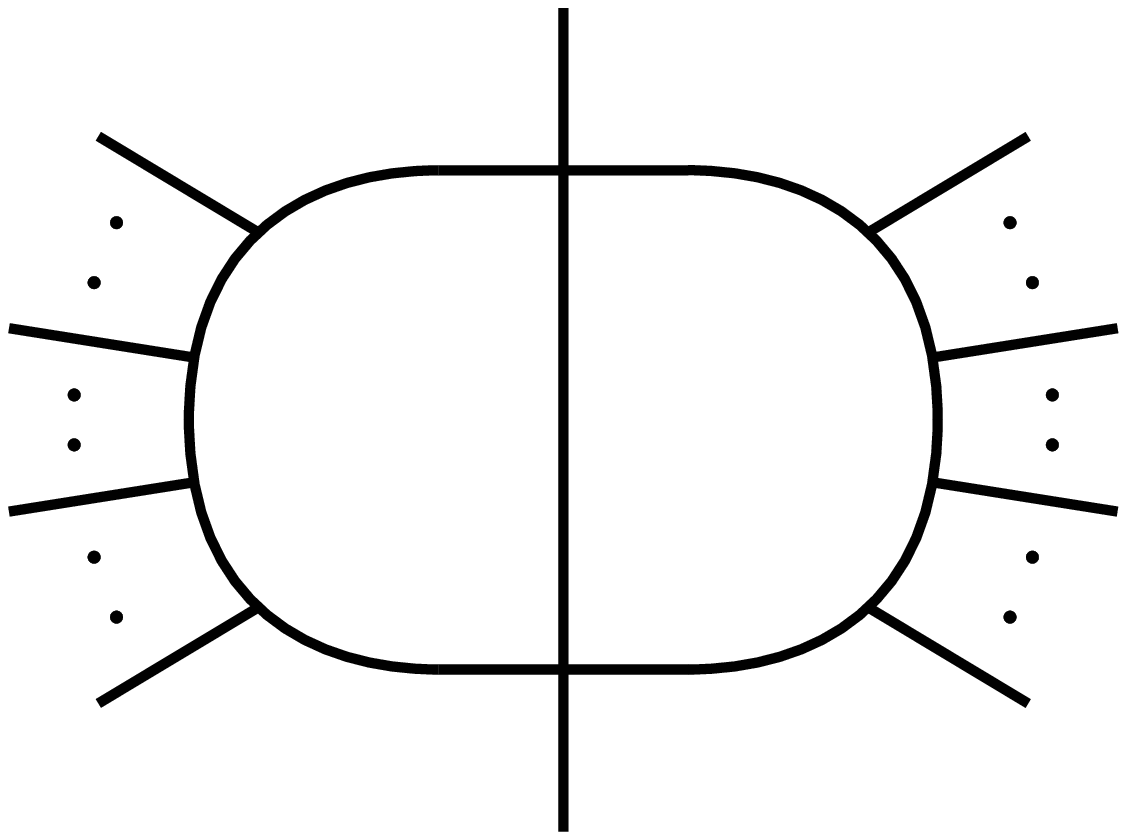}
\center{(a)}
\end{minipage}
\hspace{0.5cm}
\begin{minipage}[b]{0.4\linewidth}
\centering
\includegraphics[scale=0.5]{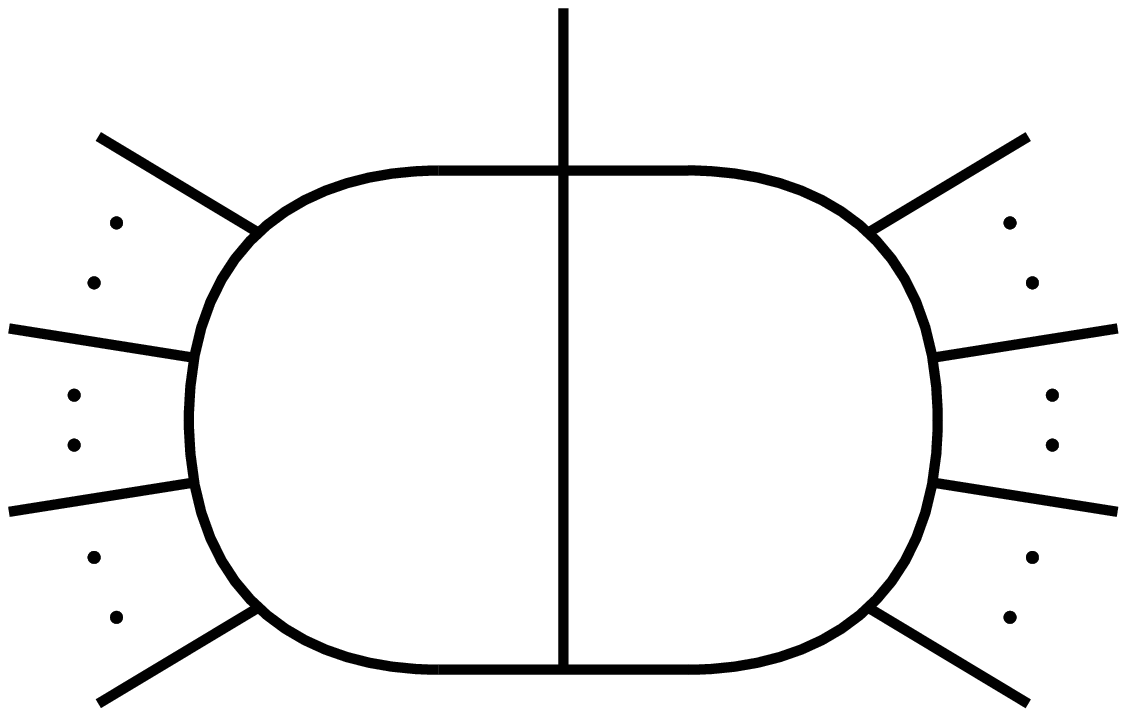}
\vspace{7mm}
\center{(b)}
\end{minipage}
\begin{minipage}[b]{0.4\linewidth}
\centering
\includegraphics[scale=0.5]{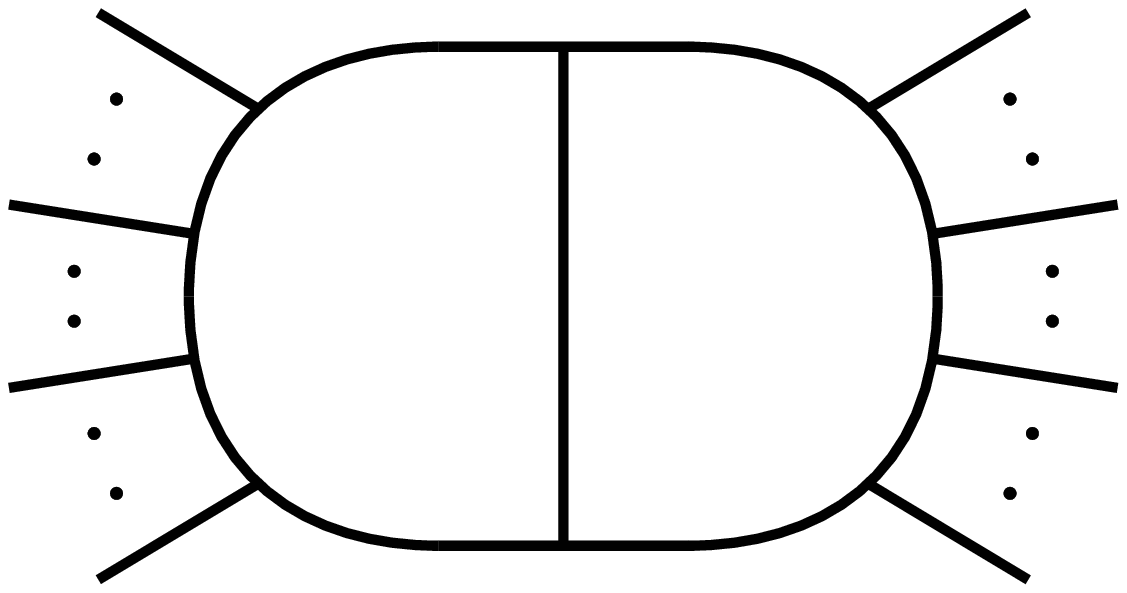}
\vspace{7mm}
\center{(c)}
\end{minipage}
 \caption{The three basic types of two-loop planar integrals, 
labeled by the number
of legs attached to each leg of the vacuum diagram: (a) $P_{n_1,n_2}$, (b)
$P^*_{n_1,n_2}$, (c) $P^{**}_{n_1,n_2}$.} 
\label{TwoLoopPlanarIntegralsFigure}
\end{figure}

We turn now to our main object of study, the planar two-loop
integrals.  We can organize the different integral skeletons we
obtain, representing only the propagators, into five classes.
\begin{figure}[ht]
\includegraphics[scale=0.5]{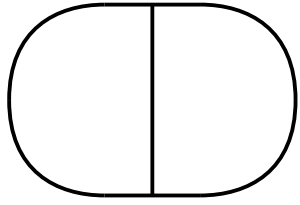}
 \caption{The non-trivial two-loop vacuum diagram.} 
\label{TwoLoopVacuumDiagramFigure}
\end{figure}
  Three classes,
 depicted
in fig.~\ref{TwoLoopPlanarIntegralsFigure}, arise from
attaching external legs to the non-trivial 
two-loop vacuum diagram shown in fig.~\ref{TwoLoopVacuumDiagramFigure}.
  We obtain planar integrals
by attaching external legs to one or two of the internal lines, and
possibly to its vertices.  Were we to attach external legs to the
third internal line as well (here, the middle line), we would obtain
non-planar integrals.  This gives rise to three of the five types of
two-loop planar integrals; the remaining types are simply products of
one-loop integrals.  We label the integrals according to the number of
external legs attached to each of the vacuum diagram's internal lines,
denoting the absence of lines attached to vertices by stars.  The three
types of integrals are,
\begin{eqnarray}
P_{n_1,n_2} &=& (-i)^2\int {d^D\ell_1\over (2\pi)^D} {d^D\ell_2\over (2\pi)^D}
\;{1\over \ell_1^2 (\ell_1-K_1)^2\cdots (\ell_1-K_{1\cdots n_1})^2(\ell_1+\ell_2+K_{n_1+n_2+2})^2}
\nn\\&&\hphantom{ \int {d^D\ell_1\over (2\pi)^D} {d^D\ell_2\over (2\pi)^D} }
  \times {1\over\ell_2^2 (\ell_2-K_{n_1+n_2+1})^2\cdots (\ell_2-K_{(n_1+2)\cdots(n_1+n_2+1)})^2}
\,,\nn\\
P^*_{n_1,n_2} &=& (-i)^2\int {d^D\ell_1\over (2\pi)^D} {d^D\ell_2\over (2\pi)^D}
\;{1\over \ell_1^2 (\ell_1-K_1)^2\cdots (\ell_1-K_{1\cdots n_1})^2(\ell_1+\ell_2)^2}
\nn\\&&\hphantom{ \int {d^D\ell_1\over (2\pi)^D} {d^D\ell_2\over (2\pi)^D} }
  \times {1\over\ell_2^2 (\ell_2-K_{n_1+n_2+1})^2\cdots (\ell_2-K_{(n_1+2)\cdots(n_1+n_2+1)})^2}
\,,\label{TwoLoopPlanarIntegrals}\\
P^{**}_{n_1,n_2} &=& (-i)^2\int {d^D\ell_1\over (2\pi)^D} {d^D\ell_2\over (2\pi)^D}
\;{1\over \ell_1^2 (\ell_1-K_1)^2\cdots (\ell_1-K_{1\cdots n_1})^2(\ell_1+\ell_2)^2}
\nn\\&&\hphantom{ \int {d^D\ell_1\over (2\pi)^D} {d^D\ell_2\over (2\pi)^D} }
  \times {1\over\ell_2^2 (\ell_2-K_{n_1+n_2})^2\cdots (\ell_2-K_{(n_1+1)\cdots(n_1+n_2)})^2}
\,,\nn\end{eqnarray}
\begin{figure}[t]
\begin{minipage}[b]{0.4\linewidth}
\centering
\includegraphics[scale=0.5]{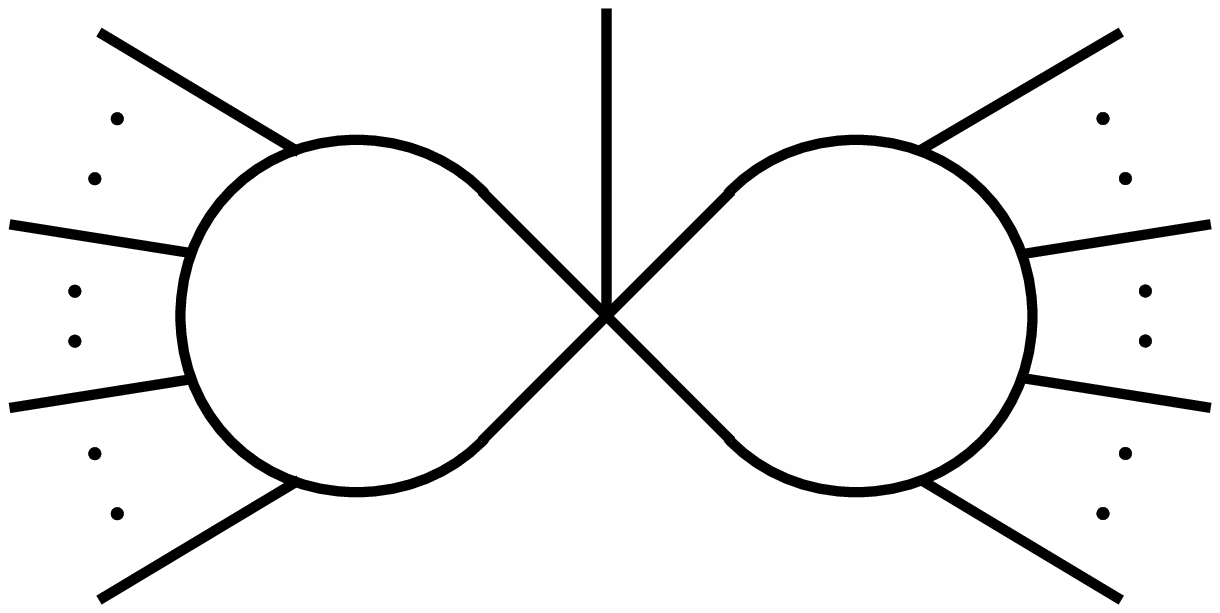}
\center{(a)}
\end{minipage}
\hspace{0.5cm}
\begin{minipage}[b]{0.4\linewidth}
\centering
\includegraphics[scale=0.5]{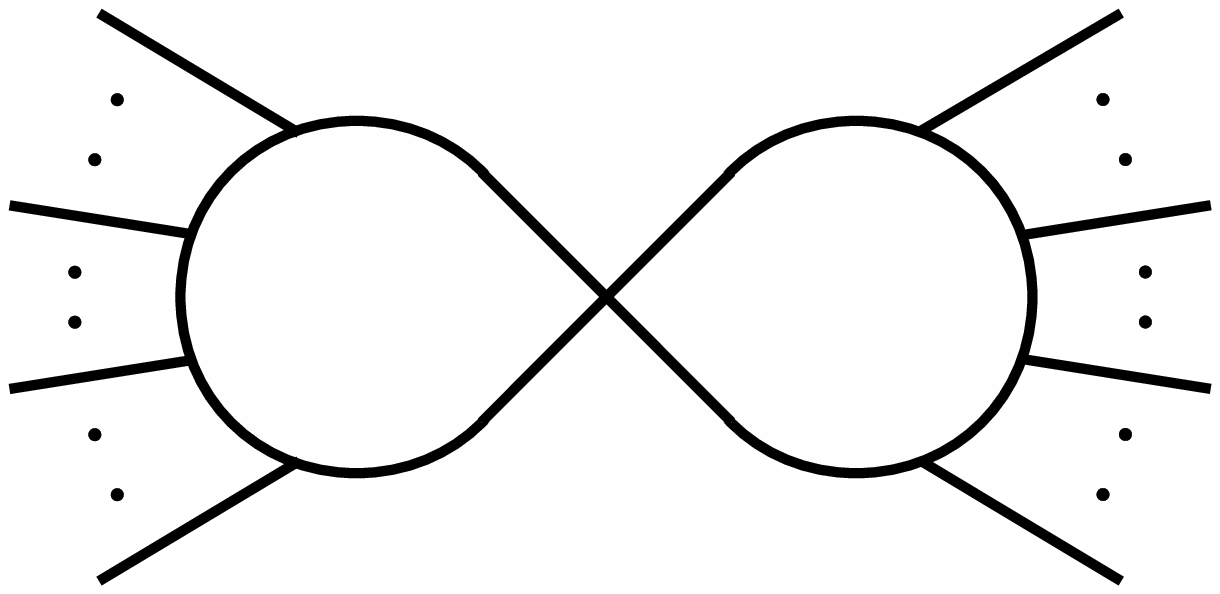}
\center{(b)}
\end{minipage}
 \caption{Two-loop integrals which are
products of one-loop integrals, 
labeled by the number
of legs attached to each leg of the vacuum diagram: (a) $\OneLoopProduct_{n_1,n_2}$, (b)
$\OneLoopProduct^*_{n_1,n_2}$.} 
\label{TwoLoopProductIntegralsFigure}
\end{figure}
along with products of two one-loop integrals
shown in fig.~\ref{TwoLoopProductIntegralsFigure},
\begin{eqnarray}
\OneLoopProduct_{n_1,n_2} &=& (-i)^2\int {d^D\ell_1\over (2\pi)^D} {d^D\ell_2\over (2\pi)^D}
\;{1\over \ell_1^2 (\ell_1-K_1)^2\cdots (\ell_1-K_{1\cdots n_1})^2}
\nn\\&&\hphantom{ \int {d^D\ell_1\over (2\pi)^D} {d^D\ell_2\over (2\pi)^D} }
  \times {1\over\ell_2^2 (\ell_2-K_{n_1+n_2+1})^2\cdots (\ell_2-K_{(n_1+2)\cdots(n_1+n_2+1)})^2}\,,\nn\\
\OneLoopProduct^*_{n_1,n_2} &=& (-i)^2\int {d^D\ell_1\over (2\pi)^D} {d^D\ell_2\over (2\pi)^D}
\;{1\over \ell_1^2 (\ell_1-K_1)^2\cdots (\ell_1-K_{1\cdots n_1})^2}
\nn\\&&\hphantom{ \int {d^D\ell_1\over (2\pi)^D} {d^D\ell_2\over (2\pi)^D} }
  \times {1\over\ell_2^2 (\ell_2-K_{n_1+n_2})^2\cdots (\ell_2-K_{(n_1+1)\cdots(n_1+n_2)})^2}
\,,
\end{eqnarray}
\def\tp{\!+\!}
so that $P_{n_1,n_2}$ is an $(n_1\tp n_2\tp 2)$-point integral, 
$P^*_{n_1,n_2}$ and $\OneLoopProduct_{n_1,n_2}$ are
$(n_1\tp n_2\tp 1)$-point integrals, and $P^{**}_{n_1,n_2}$ and $\OneLoopProduct^*_{n_1,n_2}$
are $(n_1\tp n_2)$-point integrals.
Without loss of generality, we may take $n_1\ge n_2$.

In our discussion below, we will focus on the $P_{n_1,n_2}$ integrals.
Similar arguments typically apply to the $P^{*,**}_{n_1,n_2}$.
We may also observe that,
\begin{eqnarray}
P^*_{n_1,n_2}(K_1,\ldots,K_{n_1+n_2+1}) &=& P_{n_1,n_2}(K_1,\ldots,K_{n_1+n_2+1},0)\,,\nn\\
P^{**}_{n_1,n_2}(K_1,\ldots,K_{n_1+n_2}) &=& P^*_{n_1,n_2}(K_1,\ldots,K_{n_1},0,
                                                           K_{n_1+1},\ldots,K_{n_1+n_2})\,.
\end{eqnarray}
so that the values of $P^*$ and $P^{**}$ are known in terms of $P$.
Nonetheless, their different branch cut structures strongly suggest
that the former are algebraically independent of the latter.
In explicit examples in 
sections~\ref{MasslessDoubleBoxSection}--\ref{SixPointSection}, we examine
$P^{**}$ integrals.

These integrals will arise in the leading-color contributions to
two-loop QCD amplitudes, including amplitudes for production of
electroweak bosons or other particles coupled to quarks.  Just as in
the one-loop case, all internal lines will be massless, but the
external legs of the integrals can correspond to sums of external
momenta, and hence can be either massless or massive.  Each of the
vertices can come with a power of the corresponding loop momentum, and
each of the three-point internal vertices in $P^{\natural,*,**}$ can also come
with a power of either $\ell_1$ or $\ell_2$, so that we should
consider tensor integrals with up to $(n_1+2)$ powers of $\ell_1$
along with $n_2$ powers of $\ell_2$, or alternatively
up to $(n_1+1,n_2+1)$ or $(n_1,n_2+2)$ powers of the two loop momenta
$(\ell_1,\ell_2)$.

\subsection{Reduction of High-Multiplicity Integrals with Non-Trivial Numerators}
\label{NonTrivialNumeratorReduction}

We begin our discussion of integral reduction at two loops by
considering tensor integrals with $n_1\geq 4$, $P_{n_1,n_2}[\ell\cdot
  v_1\,\ell\cdot v_2\cdots\ell\cdot v_n]$, where each $\ell$ can be
either $\ell_1$ or $\ell_2$.  We can use the expansion of
\eqn{MomentumExpansion}, with the external momenta $b_1,\ldots, b_4$
chosen amongst the first $n_1$ momenta.  This leads us to consider
integrals with numerators containing factors of $\ell_1\cdot K_j$,
where $1\leq j\leq n_1$.  As in the one-loop case, these numerators
are reducible,
\begin{equation}
\ell_1\cdot K_j = \frac12 \bigl[ (\ell_1-K_{1\cdots(j-1)})^2 - (\ell_1-K_{1\cdots j})^2 
                                + K_{1\cdots j}^2 - K_{1\cdots(j-1)}^2\bigr]\,.
\label{DotProductReductionB}
\end{equation}
The first two terms lead to integrals with smaller indices
($P_{n_1-1,n_2}$, $P^*_{n_1-1,n_2}$, or one of $P_{n_1,n_1-1}$ and
$P^*_{n_1,n_1-1}$ in the case $n_1=n_2$) and simpler tensors, while
the last two lead to integrals with simpler tensors.  Repeating this
procedure, including application to $\ell_2$,
 leads to tensor integrals $P_{n_1,n_2}$ with $n_1\leq 4$ and
$n_2<4$, or integrals with general $(n_1,n_2)$ but with trivial
numerators.

\subsection{Reduction of High-Multiplicity Integrals with Trivial Numerators}
\label{TrivialNumeratorReduction}

Once we have eliminated high-multiplicity tensor integrals, or
equivalently those with non-trivial polynomials in $\ell_{1,2}$ in the
numerator, we must consider integrals with trivial numerators but
arbitrary number of external legs.  To reduce $P_{n_1,n_2}$ integrals
with $n_1\ge 5$, we can make use of the same Gram determinant as in
the one-loop case,
\begin{equation}
\Gram{\ell_1,1,2,3,4}{5,1,2,3,4} = 0\,.
\label{VanishingGramDet}
\end{equation}
\def\indentA{\hskip -40mm}
We will obtain the reduction,
\begin{eqnarray}
P_{n_1,n_2}(K_1,\ldots,K_{n_1+n_2+2}) &=&\nn\\
&&\indentA c_1 P_{n_1-1,n_2}(K_2,\ldots,K_{(n_1+n_2+2)1})
                      +c_2 P_{n_1-1,n_2}(K_{12},K_3,\ldots,K_{n_1+n_2+2})\\
                    &&\indentA +c_3 P_{n_1-1,n_2}(K_{1},K_{23},K_4,\ldots,K_{n_1+n_2+2})
                      +c_4 P_{n_1-1,n_2}(K_{1},K_2,K_{34},K_5,\ldots,K_{n_1+n_2+2})\nn\\
                    &&\indentA  +c_5 P_{n_1-1,n_2}(K_{1},\ldots,K_{45},\ldots,K_{n_1+n_2+2})
                      +c_6 P_{n_1-1,n_2}(K_{1},\ldots,K_{56},\ldots,K_{n_1+n_2+2})\,,\nn
\end{eqnarray}
where the coefficients $c_i$ are given in
\eqn{HexagonReductionCoefficients}, the same ones as in the one-loop
reduction.  This reduction involves only propagators dependent solely
on $\ell_1$.
(In the case $n_1=n_2$,
$P_{n_1-1,n_2}(\{K_i\}_1^{n_1};\{K_j\}_{n_1+2}^{n_1+n_2+1})$ is given
by the flipped integral
$P_{n_2,n_1-1}(\{K_j\}_{n_1+n_2+1}^{n_1+2};\{K_i\}_{n_1}^1)$.)  This
reduces the set of scalar integrals to $P_{n_1,n_2}$ with $ n_2 \leq
n_1\leq 4$.  This means we have a finite (if large) set of integrals
in terms of which we can express any planar two-loop integral, and
hence any planar two-loop amplitude.

This reduction generalizes both to non-planar and to higher-loop 
integrals.  The details are in some cases more intricate, but
at two loops, we can reduce all integrals with more than eleven
propagators.  We can see this using a variety of Gram determinant
equations similar to \eqn{VanishingGramDet}.

There is another, more general, way of looking at this question.  
Let us label the momenta in the two-loop vacuum diagram of 
fig.~\ref{TwoLoopVacuumDiagramFigure} by $\ell_1$, $\ell_2$,
and $\ell_3$.  They are not independent, because $\ell_1+\ell_2+\ell_3=0$,
but can be treated symmetrically.
There can at most be eleven independent invariants $t_i$ involving
the loop momenta, namely the three
squares of the loop momenta,
\begin{equation}
\ell_1^2,\quad \ell_2^2,\quad \ell_3^2\,;
\end{equation}
and eight invariants built out of loop momenta, of the form
\begin{equation}
\ell_j\cdot k_i\,,
\end{equation}
where four $k_i$ are selected out of the external momenta.  
Because the external momenta are strictly four-dimensional, we
can express all remaining ones, and hence all invariants involving them,
in terms of these four.
We can choose the invariants to be all manifestly reducible. 
If we have more than eight external lines attached to the lines
in the vacuum diagram
(that is, excluding legs attached directly to the vertices in it),
then there are more than eleven propagators with denominators
$d_i$, and accordingly we can
write down non-trivial equations,
\begin{equation}
0 = d_i - \sum_j c_j t_j
\end{equation}
for denominators beyond the eleventh.
Inserting this equation into the
integrand allows us to reduce the integrals with
more than eleven propagators to simpler integrals, at arbitrary
$D$ or equivalently to all orders in $\eps$.  (This assumes that
the coefficient of the original integral is non-zero, which 
the earlier discussion demonstrates for the two-loop case.)

\subsection{Integration-by-Parts Without Doubled Propagators}

The reductions discussed in the previous subsections show that all
required integrals for a planar $n$-point two-loop amplitude can be
written in terms of the $P_{4,4}$ integral with a trivial numerator;
$P_{4,n_2<4}$ integrals with trivial numerators or  numerators
dependent only on $\ell_2$; and $P_{n_1<4,n_2\le n_1}$ integrals with
trivial numerators or  numerators dependent on either or both of the
loop momenta.  Of course, some numerators can still be written as a
difference of denominators, as in \eqn{DotProductReductionB}.  The
corresponding integrals can then be reduced.  The remaining
irreducible integrals, for $n_1+n_2\ge 5$, are those with irreducible
numerators, which cannot be written in such a way.  For example, in
$P_{4,3}$, $\ell_2\cdot K_1$ would be an irreducible numerator.
(Some integrals with $n_1+n_2\le 4$ require a more specialized analysis,
just as at one loop, and for the most part
 we shall not consider them in the present
article.)

To reduce these integrals, where possible, we will employ the
integration-by-parts (IBP) technique first introduced long ago by
Chetyrkin and Tkachov~\cite{IBP}, and refined into a
general-purpose algorithm by Laporta~\cite{Laporta}.

The IBP technique as outlined by Laporta, and as implemented 
in \AIR~\cite{AIR},
\FIRE{}~\cite{FIRE}, and~\Reduze{}~\cite{Reduze},
relies on writing down all
possible equations from introducing a differentiation inside the
integrand,
\begin{equation}
P_{n_1,n_2}\biggl[{\partial\over \partial\ell_j^\mu} v^\mu\biggr] =
-\int {d^D\ell_1\over (2\pi)^D} {d^D\ell_2\over (2\pi)^D}\;
{\partial\over \partial\ell_j^\mu} {v^\mu\over D(\ell_1,\ell_2,\{K_i\})} = 0\,,
\label{IBPEquation}
\end{equation}
where $D(\ell_1,\ell_2,\{K_i\})$ is the denominator found in
\eqn{TwoLoopPlanarIntegrals}.  The simplest choices for $v^\mu$ are
the set of external momenta, along with the two loop momenta.
The use of dimensional regularization ensures that there is no boundary
term in this equation.

With these choices, however, the resulting equations involve not only
the integrals of interest (as well as simpler planar integrals), but
also integrals with doubled propagators.  These arise from derivatives
hitting the denominators of the integrals.  
Moreover, such integrals can have worse infrared singularities,
so that their use results in the appearance
of additional inverse powers of $\e$ in coefficients.
This in turn would require them to be known (either
analytically or numerically) to higher orders in $\e$.
These integrals do not
arise directly in the computation of gauge-theory amplitudes, and usually we do
not wish to introduce them at the stage of solving these equations.
In all cases studied to
date, it has been possible to eliminate such integrals (at a cost of
retaining some integrals with non-trivial numerators), and it seems
plausible that this holds true generally.  
Their elimination requires the consideration of very
large systems of equations, performed using the Laporta algorithm.  In
addition to the considerable computational complexity of these
systems, which has made it difficult to proceed in explicit examples
beyond four-point integrals, it is also far from clear how to
characterize these systems in the general case.

For this reason, we seek to simplify the equations we obtain by
eliminating unwanted integrals, those with doubled propagators, from
the very start.  We will do so by making special choices of the
$v^\mu$ vectors in \eqn{IBPEquation}.  
For example, we could choose vectors whose dot
product with the numerator resulting from differentiating any
propagator vanishes,
\begin{equation}
v \cdot (\ell-K) = 0\,.
\label{vanishingDotProduct}
\end{equation}
As these expressions are the coefficients of the doubled propagators, 
this vanishing will
ensure that doubled propagators are absent.  We can construct such vectors
using Gram determinants; defining
\begin{equation}
\Gram{\mu,b_1,b_2,b_3,b_4}{d_1,d_2,d_3,d_4,d_5} \equiv
 \frac{\partial}{\partial w_\mu} \Gram{w,b_1,b_2,b_3,b_4}{d_1,d_2,d_3,d_4,d_5}\,,
\end{equation}
vectors $v$ of the form
\begin{equation}
v^\mu = \Gram{\mu,\ell_1,\ell_2,6,7,8}{\ell_2,\ell_1,1,2,3,4}
\label{vExampleI}
\end{equation}
will have the desired property with respect
to propagators depending only on $\ell_2$.  For example, the IBP equation,
\begin{equation}
P_{4,3}\bigl[\partial_2\cdot v\bigr] = 0\,
\end{equation}
(where $\partial_j = \partial/\partial\ell_j$), will be free of
doubled propagators.  Because $\ell_{1,2}$ are $D$-dimensional vectors,
Gram determinants like that in \eqn{vExampleI} give the most general
solution to \eqn{vanishingDotProduct}.

\def\Rem{\mathop{\rm Rem}\nolimits}
\def\sig{\sigma}
However, this solution is not general enough for our purposes.
The problem is that while the constraint~(\ref{vanishingDotProduct}) 
is sufficient, it is not
necessary.  It is in general too strong a constraint, 
and in practice we would miss
equations if we insisted on it.  The weaker constraint, that
\begin{equation}
v \cdot (\ell-K) \propto (\ell-K)^2\,,
\label{divisibleDotProduct}
\end{equation}
suffices to remove the doubled propagator as well.  This constraint
can be expressed as the requirement that there be no remainder upon
synthetic division of $v\cdot (\ell-K)$ by the propagator denominator
$(\ell-K)^2$.  We must impose this constraint for every propagator.  
For an integral with $n_d$ propagators, that is with a denominator in the
form,
\begin{equation}
W_n^{-1} \equiv \prod_{j=1}^{n_d} d_j
= \prod_{j=1}^{n_d} (\sig_{j1}\ell_1+\sig_{j2}\ell_2-K_j)^2\,,
\end{equation}
we must impose the $n_d$ equations
\begin{equation}
\Rem {\bigl[v_1\cdot {\partial\over \partial\ell_1} 
            +v_2\cdot {\partial\over \partial\ell_2} \bigr]
           (\sig_{j1}\ell_1+\sig_{j2}\ell_2-K_j)^2
      \over (\sig_{j1}\ell_1+\sig_{j2}\ell_2-K_j)^2} = 0\,,
\label{RemainderEquationI}
\end{equation}
or equivalently,
\begin{equation}
\Rem {\bigl[\sig_{j1} v_1+\sig_{j2} v_2\bigr]
            \cdot (\sig_{j1}\ell_1+\sig_{j2}\ell_2-K_j)
      \over (\sig_{j1}\ell_1+\sig_{j2}\ell_2-K_j)^2} = 0\,,
\label{RemainderEquationII}
\end{equation}
where $\Rem$ denotes the remainder on synthetic division (using
either $\ell_i$ as a variable).  (In these equations, the $\sig_{j}$
will be $\pm1$ or 0.)
These equations are for vectors
$v_{1,2}$ built out of the loop momenta, external momenta,
and dot products thereof.
We will discuss how to find the general solution to these
equations in the next section.  (By convention, we order the denominators
as follows: first those depending only on the first loop momentum; then
those depending only on the second loop momentum; and finally, those
depending on both loop momenta.)

Before trying to solve the equations, let us try to characterize
the solutions better.
These have several general properties that will
be helpful in finding and using these vectors.  For example,
if we have a 
pair of vectors, $\{v_1^{(0)},v_2^{(0)}\}$ that 
satisfy~\eqn{RemainderEquationI}, then any multiple of the pair
is also a solution.  In particular, multiplying by any Lorentz
invariant involving either of the two loop momenta gives us
a solution.

Not all these additional solutions are useful, however.  We can
divide these Lorentz invariants into two types: the reducible ones,
expressible as a linear combination of propagator denominators and
external invariants, and irreducible ones whose dependence on
the loop momenta cannot be expressed using propagator denominators.
While multiplying by an invariant of the former type does yield
a solution to the constraints~(\ref{RemainderEquationI}), it is
not a useful solution, because it does not yield an independent
equation for the integrals of irreducible numerators.  To see this,
let us write out the resulting IBP equations.  The original equation
is,
\begin{equation}
I\bigl[\partial_1 \cdot (v_1 W) + \partial_2\cdot (v_2 W)\bigr] = 0\,.
\end{equation}
Multiplying the vectors by a factor $f$ gives a sum of two terms,
\begin{equation}
I\bigl[f \bigl(\partial_1 \cdot (v_1 W) + \partial_2\cdot (v_2 W)\bigr)\bigr] 
+ I\bigl[ W (v_1 \cdot \partial_1 f + v_2 \cdot \partial_2 f)\bigr]= 0\,.
\label{SolutionWithFactor}
\end{equation}
The first term contains reduced integrals (that is, with fewer propagators)
and terms proportional to the original equation.  What about the second
term?  If  $f$ is reducible, its derivative can be written as a linear
combination of the derivatives of the denominators in $W$.  Because
of \eqn{RemainderEquationII}, the sum in parenthesis is strictly
reducible, that is reducible {\it without\/} adding any terms proportional 
to external invariants.  The second term in \eqn{SolutionWithFactor}
thus contributes only reduced integrals.  Accordingly, only irreducible
factors $f$ can give rise to new IBP equations.

Indeed, the solution itself won't be useful if both of the vectors
$v_1$ and $v_2$ are reducible; we will therefore restrict attention to
solutions in which at least one is irreducible, that is, at least one
term in one of the pair $v_{1,2}$ is irreducible.  
Note that not all independent vectors
will lead to independent IBP equations; but because the independence
of the IBP equations can depend on whether the
dimensional regulator $\e$ is taken to zero or
not, we leave that assessment to a later stage.


It is possible to find even weaker constraints that remove double
propagators, by adding a `total derivative', that is a function
which integrates to zero, to the right-hand side of \eqn{divisibleDotProduct}.  We will not consider such right-hand
sides.
In the examples we consider below, they are not
necessary, though we know of no general proof.

Integration-by-parts equations can be supplemented with Lorentz-invariance
equations~\cite{GehrmannRemiddi}, 
using operators built out of derivatives with respect to 
external momenta.  In general, these are not independent of the complete
tower of IBP equations~\cite{LIdependent}; because we will be able
to generate the complete tower of IBP equations, we do not need
to consider Lorentz-invariance equations.

As mentioned in the introduction, for generalized unitarity, we only need
to know the basis integrals, that is the set of integrals left
independent by the full set of IBP equations.  In order to find this set,
we could of course solve for the IBP-generating vectors analytically, and
then construct the set of IBP equations analytically as well.  However,
it suffices to solve for these vectors for a randomly-chosen (`generic')
numerical configuration of external momenta.  For higher-point integrals,
or integrals with many massive external legs, this can greatly reduce the
complexity of the calculation, and in particular the memory required
to solve for the IBP-generating vectors.

\subsection{Additional Identities to $\Ord(\e)$}
\label{AdditionalIdentitiesSubsection}

The integration-by-parts identities give relations between different
integrals that are valid for arbitrary dimension $D$, or equivalently, to all
orders in the dimensional regulator $\e$.  However, in practical 
calculations at a given order in perturbation theory
we are interested in computing terms only through
$\Ord(\e^0)$, and we are quite willing to drop terms of $\Ord(\e)$
or higher.  Additional relations between integrals, even if
they are only valid through $\Ord(\e^0)$, are for these practical
purposes just as good as relations that hold to all orders in $\e$.
The reduction of the one-loop pentagon integral, as we reviewed in
\sect{OneLoopSection}, is exactly this kind of relation.  In general,
we can write down several forms of integrands leading to integrals
of $\Ord(\e)$, built of Gram determinants or products thereof,
\begin{eqnarray}
&&\GramO{\ell_1,b_1,b_2,b_3,b_4}\GramO{\ell_2,b'_1,b'_2,b'_3,b'_4}\,;\quad
\Gram{\ell_1,b_1,b_2,b_3,b_4}{\ell_2,b'_1,b'_2,b'_3,b'_4}\,;\nn\\
&&\Gram{\ell_1,\ell_2,b_1,b_2,b_3}{\ell_1,\ell_2,b''_1,b''_2,b''_3}\,;\quad
\Gram{\ell_1,\ell_2,b_1,b_2,b_3,b_4}{\ell_1,\ell_2,b'_1,b'_2,b'_3,b'_4}\,.
\label{EpsGramDets}
\end{eqnarray}
where the momenta attached to the first loop (through which $\ell_1$
flows) are contained either within the set $\{b_1,b_2,b_3\}$ or
within the set $\{b''_1,b''_2,b''_3\}$, and the momenta attached
to the second loop are contained within the other of the two sets.
These Gram determinants all vanish when either loop momentum approaches a
potential (on-shell) collinear or soft configuration, thereby removing
the corresponding divergences from the integral, and rendering it finite.
In addition, the Gram determinants vanish when both loop momenta are
four-dimensional, so that the integrals are of $\Ord(\e)$.  We can 
also write down differences of expressions yielding
finite integrals which will again vanish when both loop momenta are
four-dimensional, so that the resulting integrals are again of
$\Ord(\e)$,
\begin{equation}
\Gram{\ell_1,b_1,b_2,b_3}{\ell_1,b_4,b_5,b_6}
\Gram{\ell_2,b'_1,b'_2,b'_3}{\ell_2,b'_4,b'_5,b'_6}
-\Gram{\ell_1,b_1,b_2,b_3}{\ell_2,b'_4,b'_5,b'_6}
\Gram{\ell_2,b'_1,b'_2,b'_3}{\ell_1,b_4,b_5,b_6}
\label{EpsGramDetsII}
\end{equation}
where the legs attached to the first loop are all represented amongst
the $b_i$, and the legs attached to the second loop, amongst the $b'_i$.
(For $P^*$ and $P^{**}$ integrals, $k_1$ must also be amongst the $b'_i$,
and $k_n$ amongst the $b_i$; for $P^{**}$, $k_{n_1+1}$ must be amongst
the $b_i$, and $k_{n_1}$ amongst the $b'_i$.)

Not all
of these determinants will necessarily lead to useful equations reducing the 
basis.  We can also consider integrals with numerators containing
a product of one of these Gram determinants and other irreducible factors,
so long as the integrals are ultraviolet-finite (which can be determined
by power-counting).
As is true for the IBP generating vectors, we can also generate
additional identities for a randomly-chosen configuration of external
momenta; this will be sufficient to identify the integrals that are
independent through $\Ord(\eps^0)$.  Typically, we will first solve
all $D$-dimensional IBP equations, and use the solutions of those equations
(in analytical or numerical form) 
to reduce the integrals obtained from inserting
Gram determinants into the numerator; this will provide
additional identities to $\Ord(\eps^0)$ between the independent master
integrals.

We can write down additional Gram determinants beyond 
those given in \eqn{EpsGramDets}, 
\begin{equation}
\Gram{\ell_1,\ell_2,b_1,b_2,b_3}{\ell_1,b'_1,b'_2,b'_3,b'_4}\,;
\Gram{\ell_1,\ell_2,b_1,b_2,b_3}{\ell_2,b'_1,b'_2,b'_3,b'_4}\,,
\label{EpsGramDetsIII}
\end{equation}
where all momenta attached to the second loop 
(with loop momentum $\ell_2$) are represented amongst
$\{b_1,b_2,b_3\}$ in the first case, and similarly for the momenta
attached to the first
loop in the second case.
However, these determinants give rise to integrands which are
in fact total derivatives, and hence the corresponding integrals vanish
identically.  To see this, consider the following vector,
\begin{equation}
\Gram{\mu,\ell_1,\ell_2,b_1,b_2,b_3}{\ell_1,\ell_2,b'_1,b'_2,b'_3,b'_4}\,,
\end{equation}
where all momenta attached to loop in which $\ell_2$ flows are in
$\{b_{1,2,3}\}$.
The vector's dot product with the derivative of any propagator with 
respect to
$\ell_2$ will vanish so that only the derivative of the Gram determinant
itself can enter any equation; but that derivative is proportional to the
first determinant in \eqn{EpsGramDetsIII}.
Accordingly, we do not need to consider the forms in
\eqn{EpsGramDetsIII} if we have already solved
the IBP equations.  If we include additional irreducible prefactors,
we will again either obtain an expression proportional to the determinants
in \eqn{EpsGramDetsIII} or to linear combinations of them and
the last determinant in \eqn{EpsGramDets}.  Equations similar to 
those considered in this section were obtained for six-point
integrals by Cachazo, Spradlin,
and Volovich~\cite{LeadingSingularityCalculation} using leading singularities.

\section{IBP-Generating Vectors}
\label{IBPGeneratingSection}

In order to find the general form of vectors leading to IBP equations
free of doubled propagators, we must find
the general solution to
the set of equations~(\ref{RemainderEquationII}).  We begin by rewriting
them in a somewhat more convenient form,
\begin{equation}
\bigl[\sig_{j1} v_1+\sig_{j2} v_2\bigr]
  \cdot (\sig_{j1}\ell_1+\sig_{j2}\ell_2-K_j)
+ u_j \,(\sig_{j1}\ell_1+\sig_{j2}\ell_2-K_j)^2 = 0\,,
\label{IBPBasicVectorEquation}
\end{equation}
where $u_j$ is a polynomial in the various independent Lorentz invariants of
the loop and external momenta.  Because the different propagator denominators
are independent (the integrals for which this is not true we have already
treated in sects.~\ref{NonTrivialNumeratorReduction} 
and~\ref{TrivialNumeratorReduction}), this equation must hold for
each of the $n_d$ propagators independently.

Let us also write a general form for
the $v_i^\mu$,
\def\coeffA#1{c^{(#1)}}
\begin{equation}
v_i^\mu = \coeffA{\ell_1}_i \ell_1^\mu
+\coeffA{\ell_2}_i \ell_2^\mu +\sum_{b\in B} \coeffA{b}_i b^\mu\,,
\label{IBPVectorGeneralForm}
\end{equation}
where the sum runs over a set of $n_B$ --- up to four --- 
basis vectors for the external
momenta.  (There would be four basis vectors for integrals with five
or more external legs, and $n-1$ vectors for integrals with fewer.)
Each of the coefficients $\coeffA{x}_i$ is again a polynomial in
the various independent Lorentz invariants.  

We consider as independent variables only invariants that are independent
with respect to the loop momenta.  That is, $\ell_1^2$ and $\ell_2^2$
are independent, as are each of these with respect to $\ell_1\cdot k_1$,
and a given invariant of the external momenta, say $k_1\cdot k_2$.
However, different invariants of external momenta
are not independent, which is to say their
ratio should be treated as a constant parameter.  Let us pick the 
one independent invariant to be $s_{12} = (k_1+k_2)^2$, and define
the ratios,
\begin{eqnarray}
\chi_{ij} &&= \frac{s_{ij}}{s_{12}}\,,\nonumber\\
\chi_{i\cdots j} &&= \frac{s_{i\cdots j}}{s_{12}}\,,
\label{Parameters}\\
\mu_i &&= \frac{m_i^2}{s_{12}}\,,\nonumber
\end{eqnarray}
in order to express the remaining invariants in terms of $s_{12}$.  
(For certain integrals with fewer than four external legs, we should
pick a different invariant.)
We will term these quantities parameters.
Each
of the coefficients $\coeffA{x}_i$ would have an expression in terms
of the invariants,
\begin{equation}
V = \{
\ell_1^2\,, \ell_1\cdot \ell_2, \ell_2^2\,,
\{\ell_1\cdot b\}_{b\in B}\,,
\{\ell_2\cdot b\}_{b\in B}, s_{12}\}\,.
\label{BasicVariables}
\end{equation}
We treat these invariants as the basic symbols or variables out of
which we build the solutions.
 For example, coefficients of engineering dimension two could be
expressed as follows,
\begin{eqnarray}
\coeffA{p}_i &=&
\coeffA{p}_{i,1} s_{12}
+\sum_{b\in B}\coeffA{p}_{i,b1}\ell_1\cdot b
+\sum_{b\in B}\coeffA{p}_{i,b2}\ell_2\cdot b
+\coeffA{p}_{i,2}\ell_1^2
+\coeffA{p}_{i,3}\ell_1\cdot\ell_2
+\coeffA{p}_{i,4}\ell_2^2\,.
\end{eqnarray}
The coefficients $\coeffA{p}_{i,j}$ of each term are rational functions
of the parameters $\chi_{i\cdots j}$ and $\mu_i$.
(In order to distinguish the
different dimensions to which we will refer below, we refer to the
engineering or energy dimension of $\ell_i$ and $b$ as such, dropping
the `engineering' qualifier only when context makes it unnecessary.)

Our discussion generalizes in a straightforward way 
both to higher loops and to non-planar integrals.  At higher loops,
we will have a vector $v_l$ for each loop; the 
expansion~(\ref{IBPVectorGeneralForm}) will have a sum over all loop momenta;
and the set of variables $V$ in \eqn{BasicVariables}
will include all squares of loop momenta,
all dot products of loop momenta with each other, and all dot products
of the loop momenta with the basis vectors in $B$.  In general, some
dot products of loop momenta with other loop momenta will be irreducible,
but this does not change the derivation of the equations.  For non-planar
integrals at two loops, we will have more than a single equation involving two 
different vectors.  (This will anyway be true at higher loops.)
The general approach to solving the equations we now
outline will also carry over, though
the specific procedure
that solves the equations most efficiently in these more general 
settings remains to be investigated.  Internal masses will introduce
additional parameters in \eqn{Parameters} while leaving the
basic invariants~(\ref{BasicVariables}) and the general structure
unchanged.

Without loss of generality, we consider only solutions $v_{1,2}$
of homogeneous engineering dimension.  
We could in principle proceed by writing
down a general form for coefficients of a given engineering
dimension, starting
with dimension zero, and proceeding by increments of two.  Plugging
in these general forms into \eqn{IBPVectorGeneralForm}, and requiring
that the coefficient of each monomial in the basic 
variables~(\ref{BasicVariables}) vanish independently, we would obtain
the solutions of the given dimension.

This method of solution works quite well for finding solutions of low
dimension, but becomes very memory-intensive for higher dimensions.
Furthermore, it does not allow us to determine when we have found the
complete independent basis set of solutions, namely the set of
solutions $v_{1,2}$ to \eqn{IBPVectorGeneralForm} in terms of which
all others can written as linear combinations, with coefficients that
are polynomials in the basic variables~(\ref{BasicVariables}) and
rational in the parameters~(\ref{Parameters}).

\def\Irred{\mathop{\rm Irred}\nolimits}
As a reminder,
we are interested only in solutions for which at least one coefficient
in either $v_1$ or $v_2$ is irreducible with respect to the 
set of denominators.
Let us denote the operation of removing terms proportional to
a propagator denominators (that is, 
reducing by the set of propagator denominators)
by the operator $\Irred$.  It leaves behind only the irreducible part of
an expression.  (This operation is most naturally implemented using
a Gr\"obner basis for the propagator denominators, but as these are
all linear in the basic variables~(\ref{BasicVariables}), the use
of such a basis is not essential.)  We defer a precise definition to
later in this section.

\def\coeffB#1{\tilde c_{#1}}
We can assemble the set of equations~\eqn{IBPBasicVectorEquation} into
a single matrix equation.  To do, first assemble the various coefficients
using the relabeling,
\begin{eqnarray}
\rho(\ell_1,1) &=& 1\,,\nn\\
\rho(\ell_2,1) &=& 2\,,\nn\\
\rho(j,1) &=& j+2\,,\quad  j \in \{1,\ldots,n_B\}\,,\nn\\
\rho(\ell_1,2) &=& n_B+3\,,\label{AssemblyMapping}\\
\rho(\ell_2,2) &=& n_B+4\,,\nn\\
\rho(j,2) &=& n_B+j+4\,,j \in \{1,\ldots,n_B\}\,;\nn
\end{eqnarray}
and the definitions,
\begin{eqnarray}
\coeffB{\rho(q,i)} &=& \coeffA{q}_i\,,\quad q\in \{\ell_1,\ell_2\}\cup B\,, 
            i\in\{1,2\}\,,\nn\\
\coeffB{2n_B+4+j} &=& u_j\,,\quad j \in \{1,\ldots,n_d\}\,.
\end{eqnarray}

\def\Syz{\mathop{\rm Syz}\nolimits}
Treating the coefficients $\coeffB{}$ as a row 
vector\footnote{Row vectors provide a more natural interpretation,
as this choice also leads to treating the derivatives of the equation
with respect to this vector's
entries as row vectors, which in turn leads to a more natural
implementation in a symbolic algebra language such as {\sl Mathematica\/}.},
define
\begin{eqnarray}
E_{\rho(q,i),j} = q\cdot \partial_i d_j\,,\nn\\
E_{2n_B+4+j,j} = -d_j\,.
\label{EDefinition}
\end{eqnarray}
$E$ is a $(2n_B+4+n_d)\times n_d$-dimensional matrix; the
number of rows we will label $n_r$.  Each column corresponds to
\eqn{IBPBasicVectorEquation} for a different propagator.
We then have the following matrix equation,
\begin{equation}
\coeffB{} E = 0\,.
\label{IBPMatrixEquation}
\end{equation}
Mathematicians call each solution to this equation a syzygy of $E$.

In intermediate stages, we may need to solve not only 
homogeneous equations such as this, but also inhomogeneous
equations,
\begin{equation}
\coeffB{} E = f\,,
\label{InhomogeneousEquation}
\end{equation}
where the row vector
$f$ is independent of $\coeffB{}$ though it may depend on other 
parameters.

\def\bolde{\boldsymbol{e}}
Both of these equations are linear polynomial diophantine equations.
In the 
adiatretofluous
language of mathematicians, the former
is an equation for the syzygies of the ideal submodule 
of $Q(\{\chi_{i\cdots j},\mu_i\})[V]^{n_d}$ generated by the rows of
$E$,
\begin{equation}
\Syz(\left\langle \bolde_1,\ldots,\bolde_n\right\rangle)\,.
\end{equation}  
More
precisely, we seek a linearly-independent basis for the irreducible
elements of the syzygy module,
\begin{equation}
\Syz(\left\langle \bolde_1,\ldots,\bolde_n\right\rangle)/
\Syz_{\rm Red}(\left\langle \bolde_1,\ldots,\bolde_n\right\rangle)\,,
\end{equation}
where the ${}_{\rm Red}$ subscript denotes the reducible subspace with
respect to $\Irred$.
  In this
language, it is basically a textbook problem, though
there are aspects which require a bit more work than a textbook solution.

The solution relies on the use of Gr\"obner bases~\cite{Buchberger}.  
The reader may find an explanation of the varied uses of Gr\"obner bases, 
as well as algorithms for their construction, and the required background
material, in several textbooks~\cite{AdamsLoustaunau,GroebnerTextBooks}.
Of these, we shall primarily make use of that 
Adams and~Loustaunau~\cite{AdamsLoustaunau}.
Sturmfels gave a brief overview of Gr\"obner bases~\cite{GroebnerIntro},
and 
Lin et al.{}~\cite{LXW} also provide a nice introduction to Gr\"obner bases of
modules from a physicist's point of view.  Gr\"obner bases
have been studied for use in integral reductions
by Smirnov and~Smirnov~\cite{GroebnerIntegrals},
and have been used by Smirnov in~\FIRE{}~\cite{FIRE}, as well as in other
studies of integral reductions~\cite{Gerdt}.

Our review here will mention only the minimal material needed 
for the description of the solution.  Gr\"obner bases, amongst
other uses, provide a certain generalization of linearly-independent
bases to nonlinear multivariate polynomials.  The basic setting is
that of polynomials in a set of symbols.  In our context, the symbols
are those in the set $V$~(\ref{BasicVariables}); the coefficients
are arbitrary rational functions of the $\chi_{i\cdots j}$ and
$\mu_i$, forming the field $Q(\{\chi_{i\cdots j},\mu_i\})$ in
mathematicians' language. We will need to consider vectors
or tuples of polynomials as well as polynomials themselves.

\def\lm{\mathop{\rm lm}\nolimits}
The basic machinery requires us to choose an ordering of the
terms built out of the basic symbols, as well as of tuples of terms.  
There are various
ways of doing this, of which a lexicographic ordering is conceptually
the simplest.  While the choice of ordering will not change the space
of solutions we find, the efficiency of the (standard) algorithms we
employ {\it will\/} depend greatly on this choice.  We choose the so-called
degree-reverse lexicographic order (DRL or grevlex) for the basic
symbols, and a term-over-position (ToP) ordering for tuples of polynomials.
A generic term in a polynomial built out of the symbols $x_i$ has
the form,
\begin{equation}
c \boldsymbol{x}^{\boldsymbol{p}} \equiv c x_1^{p_1} \cdots x_n^{p_n}\,,
\end{equation}
while a generic $n$-tuple of polynomials is a sum of terms of the form,
\begin{equation}
c \boldsymbol{v} \equiv 
c \boldsymbol{x_v}\bolde_v = c x_1^{p_1} \cdots x_n^{p_n} \bolde_v\,,
\end{equation}
when the unit basis tuples are the set $\{\bolde_j\}$.  The DRL
ordering for polynomial terms starts with
 a basic ordering of the symbols $x_n,\ldots,x_1$
and orders $\boldsymbol{x}^{\boldsymbol{p}}$
before $\boldsymbol{x}^{\boldsymbol{q}}$ (denoted 
$\boldsymbol{x}^{\boldsymbol{p}}\prec\boldsymbol{x}^{\boldsymbol{q}}$)
iff,
\begin{equation}
\sum_i p_i < \sum_i q_i\quad
{\rm or}\quad
\sum_i p_i = \sum_i q_i \hbox{\rm\ and the rightmost non-zero entry in
$\boldsymbol{p}-\boldsymbol{q}$ is positive}
\end{equation}
The ToP ordering orders tuples containing a lone monomial as follows,
\begin{equation}
\boldsymbol{v} \prec \boldsymbol{u} \Leftrightarrow 
\boldsymbol{x_v} \prec \boldsymbol{x_u} \quad{\rm or}\quad
\boldsymbol{x_v} = \boldsymbol{x_u}{\rm\ and\ }
\bolde_v \prec \bolde_u\,.
\end{equation}
(The basis vectors $\bolde_j$ are ordered by their first non-zero
component.)
The {\it leading monomial\/} of a polynomial (or tuple) $P$, denoted $\lm(P)$,
is the monomial $\boldsymbol{v}$ which is last in the ordering,
$\boldsymbol{v} \succ \boldsymbol{v}'$ for all monomials $\boldsymbol{v}'$ in P
(stripped of any coefficient $c$).

\def\llongrightarrow{%
\relbar\mskip-0.5mu\joinrel\mskip-0.5mu\relbar\mskip-0.5mu\joinrel\longrightarrow}
\def\inlimit^#1{\buildrel#1\over\longrightarrow}
With an ordering chosen, we can define a polynomial reduction 
algorithm, essentially a repeated synthetic division with respect
to a basis set $B$ of polynomials (or tuples), yielding a set of 
coefficients $c$, and a remainder $r$,
\begin{equation}
p \inlimit^{B} r\,,
\end{equation}
where 
\begin{equation}
p = \sum_{b\in B} c_b b + r\,.
\label{PolynomialReduction}
\end{equation}
The coefficients are again polynomials in the basic symbols, and the
remainder is a polynomial (or tuple of polynomials if $p$ is a tuple).
At each stage of the synthetic division, a polynomial is divisible by
a selected divisor iff its leading monomial is divisible by the
divisor's leading monomial.  For the purposes of synthetic division,
we can treat $n_d$-dimensional tuples of polynomials by taking their
dot product with an $n_d$-tuple of dummy or ``tag'' variables
$(t_1,\ldots,t_{n_d})$, and then performing ordinary synthetic division
with the set of variables extended to include the tag variables $t_i$.
The ToP/DRL ordering is then given by a DRL ordering, with the tag
variables ordered before the other variables.  One must ensure that
other algorithms used maintain the linearity in these tag variables.
At the end of a calculation, the tuples can be recovered by
differentiating with respect to them.  

In general, the reduction coefficients $c_b$ in \eqn{PolynomialReduction}
are not universally defined; they depend not only on the ordering
chosen for monomials, but also on the order in which the polynomials
are taken during synthetic division.  Only if the set $B$ of polynomials
is a Gr\"obner basis will the reduction coefficients be independent
of the order in which the polynomials are taken.

\def\GB{\mathop{\rm GB}\nolimits}
Let us write out the method of solution we have used, 
reverting to physicists' language, and postponing until
the next section an explicit example, that of the massless double box.
We use the Buchberger algorithm to compute the required Gr\"obner bases,
though more sophisticated algorithms~\cite{Faugere} are available and would be
worth investigating.  Using the algorithm, we compute the Gr\"obner basis
of the set of rows of $E$, treated as $n_d$-tuples.  
We assemble the elements of the basis, again $n_d$-tuples, into a matrix
$G$.  The number of rows is determined by the number of tuples $n_g$ in
the Gr\"obner basis, 
which may be smaller, equal to, or larger than the number of
original vectors $n_r$ (which is equal to $2 n_B+n_d+4$ in the case of $E$).
In addition to the basis itself, we will need the
cofactor matrix $C$, which expresses the basis elements in terms of
the original vectors,
\begin{equation}
G = C E\,.
\end{equation}
It may be computed as a by-product of Buchberger's algorithm 
(or other algorithms) for computing the Gr\"obner basis.  Because
$G$ is a Gr\"obner basis, we may also express each original vector 
as a linear combination of the basis vectors; this defines another
matrix $Q$,
\begin{equation}
E = Q G\,.
\end{equation}

\def\boldg{\boldsymbol{g}}
\def\lcm{\mathop{\rm lcm}}
\def\boldsig{\boldsymbol{\sigma}}
In order to find the syzygies of $E$, we must first write down those
of $G$.  The syzygies of the latter --- that is, $n_g$-tuples $s$ such that
$s G = 0$ --- can be constructed as described in the 
textbooks~\cite{AdamsLoustaunau,GroebnerTextBooks}. 
The construction starts with the $S$-polynomial of two rows
or basis $n_d$-tuples $\boldsymbol{g}_{i,j}$,
\begin{equation}
S_p(\boldg_i,\boldg_j) = 
  {\lcm(\lm(\boldg_i),\lm(\boldg_j))\over\lm(\boldg_i)} \boldg_i
  -{\lcm(\lm(\boldg_i),\lm(\boldg_j))\over\lm(\boldg_j)} \boldg_j\,,
\end{equation}
where $\lcm$ denotes the least common multiple, with the added
definition that $\lcm(\bolde_i,\bolde_j) \equiv 0$ if $i\neq j$.
(The factors in front of $\boldg_{i,j}$ are then pure polynomial
terms, with the basis vectors $\bolde$ canceling out.)
The $S$-polynomial also plays a central role in the Buchberger
algorithm itself.    By construction,
this $S$-polynomial can be completely reduced over the Gr\"obner basis,
\begin{equation}
S_p(\boldg_i,\boldg_j) = \sum_{k=1}^{n_g} h^{ij}_k \boldg_k\,;
\end{equation}
Each syzygy or linear relation between the
$n_g$ Gr\"obner basis elements can be represented by an $n_g$-tuple,
with basis elements $\boldsig_1,\ldots,\boldsig_{n_g}$,
\begin{equation}
\boldsig_k G = \boldg_k\,.
\end{equation}
  We can define a basic syzygy or linear relation,
\begin{equation}
\hat s_{ij} = 
  {\lcm(\lm(\boldg_i),\lm(\boldg_j))\over\lm(\boldg_i)} \boldsig_i
  -{\lcm(\lm(\boldg_i),\lm(\boldg_j))\over\lm(\boldg_j)} \boldsig_j
 -\sum_{k=1}^{n_g} h^{ij}_k\boldsig_k\,.
\label{BasicSyzygy}
\end{equation}
The complete set of syzygies of $G$ is then generated by the set
\begin{equation}
\bigl\{ \hat s_{ij} \,\big|\, 1\leq i < j \leq n_g\bigr\}\,.
\label{SyzygyGenerators}
\end{equation}
These syzygies, treated as row vectors,
 can again be assembled into a matrix $S$.
With
$S$ in hand, the rows of $S C$ are syzygies of $E$.  In addition, because
we must have
\begin{equation}
E = Q G = Q C E\,,
\end{equation}
then $(I-Q C)E = 0$, where $I$ is the identity matrix.
The rows of $I-Q C$ are thus also syzygies of
$E$.  We extend $S C$ to add these rows.
In our application, these turn out to be relevant only
in some variants of the solution algorithms described below.

\def\Se{{\overline S}}
Not all syzygies, that is rows of $S$, are linearly independent.  
Indeed, there are typically dozens or even hundreds of syzygies.
The set of syzygies can be reduced in a variety of ways,
of which the two principal ones we use are polynomial reduction
and numerically-assisted row reduction.
 Furthermore, as
discussed earlier, we
are interested only in rows of $S C$ which are independent after
reduction with respect to the propagator denominators, and this
allows for additional reductions in number.  For this purpose,
we use the $\Irred$ operator described earlier.  We define it
as the polynomial reduction (element-by-element) with respect to
a Gr\"obner basis $G_D$ of the propagator denominators, built
over the symbols in $V$ (here using a plain lexicographic
ordering),
\begin{equation}
p \inlimit^{G_D} \Irred p\,.
\end{equation}
We apply polynomial reduction to the rows of $S$, treated as
$n_d$-tuples, using a special ordering of the underlying variables,
called the Schreyer ordering~\cite{AdamsLoustaunau,Schreyer}.  It
is defined by,
\begin{equation}
\boldsymbol{x^p} \bolde_i \prec \boldsymbol{y^q} \bolde_j
\Longleftrightarrow 
\lm(\boldsymbol{x^p} \boldg_i) \prec
\lm(\boldsymbol{y^q} \boldg_j)
\quad {\rm or}\quad
\lm(\boldsymbol{x^p} \boldg_i) =
\lm(\boldsymbol{y^q} \boldg_j) {\rm\ and\ } j < i\,.
\end{equation}
It is useful because the syzygy generators~(\ref{SyzygyGenerators})
 form a Gr\"obner basis with respect to this ordering.  When reducing
syzygies, we start with those of lowest engineering dimension,
removing those which reduce to zero (and hence are linear combinations
of other syzygies), and proceed incrementally in the engineering dimension.

\def\Vec{\mathop{\rm Vec}\nolimits}
\def\red{{\rm red}}
\def\irred{{\rm irred}}
To find the set of fully-independent solutions modulo reducibility,
we can proceed as follows to convert it to a linear algebra problem.  
We form the irreducible part of the solutions,
\begin{equation}
S^\irred = \Irred S\,,
\end{equation}
and convert them into ``tagged''polynomials using ``tag''variables as
described earlier.  We now construct a vector space, in which each
coordinate corresponds to a different monomial, and where each
monomial that may appear in any of the tagged polynomials, or in any
product of an irreducible polynomial times a tagged polynomial, is
assigned a coordinate.  Each tagged polynomial $P$ (that is, each
solution $s$) may then be mapped to a vector $\Vec(s)$, whose entries
are rational functions of the $\chi_{ij}$ and the $\mu_i$.
Independence can then be determined by linear algebra ({\it e.g.\/}
row reduction).  We can check it numerically, by evaluating the
solution for a given numerical choice of external momenta.  For a
given solution $s$, we also need to generate the vectors corresponding
to multiples of $s$ by a factor $x$ built out of the variables in $V$.
We can do this either by mapping the multiple, $\Vec(x s)$, or by
multiplying $\Vec(s)$ by the appropriate matrix.  After removing
linearly-dependent solutions, we usually end up with only a handful of
independent syzygies, which we assemble into a matrix $\Se$.

The general solution to \eqn{IBPMatrixEquation} can then be
written as follows,
\begin{equation}
\coeffB{} = (p_1\,\ldots\,p_{n_s}) \Se\,,
\end{equation}
where $n_s$ is the number of independent solutions, that is the
rows of $\Se$; and where each $p_i$ is an arbitrary polynomial
in the variables in $V$.  Because the Gr\"obner basis is finite,
and of finite engineering dimension, the complete set of
solutions is generated by a finite and finite-dimensional set,
and any effectively reducible integral will be reduced by an IBP equation
built using one of the basis
elements (with a possible irreducible prefactor determined by
the dimension of the numerator in the integral).

In order to solve inhomogeneous equations such as \eqn{InhomogeneousEquation},
which will arise in some of the variants of the solution algorithm 
presented below, we must first reduce the
right-hand side $f$ over the Gr\"obner basis of $E$,
\begin{equation}
f = q_f G + r_f\,.
\label{ReduceOverGroebner}
\end{equation}
If $r_f$ is non-vanishing, the equation has no solution.
In our context, $r_f$ will typically have one or more free coefficients
$p_i$ 
(arbitrary polynomials), and
we will need to impose additional constraints on them to ensure
that $r_f$ becomes reducible over $G$.  (It is not necessary to
make it vanish strictly.)  We can do so by solving the homogeneous
equation,
\begin{equation}
r_f - \tilde g G = 0\,,
\end{equation}
where $\tilde g$ is an $n_g$-tuple of dummy coefficients.  The solution
will express the free coefficients $p_i$ in $r_f$ in terms of a more constrained
set $p'_i$.  The expression in terms of $p'_i$ will now be reducible
over $G$.
Once we have solved this subsidiary equation 
(or if $r_f$ vanishes to begin with), a particular
solution to the inhomogeneous equation is given by
\begin{equation}
\coeffB{} = q_f C\,,
\end{equation}
because then $q_f C E = q_f G = f$.  In our solutions of inhomogeneous
equations, we will not be interested in the general solution, but
it can be obtained by adding an arbitrary solution to the corresponding
homogeneous equation with $f$ set to zero (obtained following the
steps discussed above).

\subsection{A Simple Algorithm}

The steps described in the previous section can be summarized
in a simple algorithm (Algorithm I), 
which starts as input with a matrix $E$ as in
the form~(\ref{IBPMatrixEquation}),
\begin{itemize}
\item[1.] Compute the Gr\"obner basis $G$ and the cofactor matrix $C$
for the set of $n_d$-tuples given by the rows of $E$;

\item[2.] Build the set of syzygies $S$ of $G$ using \eqn{BasicSyzygy};

\item[3.] Reduce the set of syzygies by synthetic division with
respect to previous retained syzygies, discarding those with no remainder.
It is best to proceed incrementally in the syzygies' engineering dimension;

\item[4.] Construct the matrix $Q$ which expresses $E$ in terms of $G$;

\item[5.] Construct the set of solutions, $S C$ along with $I-Q C$;

\item[6.] Reduce to a set of independent solutions with respect
to reduction by the set of propagator denominators.
\end{itemize}

This algorithm works nicely and quickly for simple cases, such as
the massless double box discussed in more detail in the next section.
However, it suffers from
very memory-intensive (and slow) intermediate stages
for more complicated cases such as the four-mass double box or
the pentabox, when the number of $\chi$ and $\mu$ parameters grows.
There is room for improvement, because a great deal of unnecessary 
information (pertaining to fully-reducible solutions to the equations) 
is computed in intermediate stages.

\subsection{An Improved Algorithm}

For these reasons, we use a somewhat more involved procedure.
The greater complexity of the procedure is balanced by simpler
execution at each stage.  The basic idea is to split up the
solution into several stages.  We can split the matrix $E$
and the desired coefficients $\coeffB{}$ into reducible
and irreducible parts,
\begin{eqnarray}
E^\irred &=& \Irred E\,,\nn\\
\coeffB{}^\irred &=& \Irred \coeffB{}\,,\\
E^\red &=& E - E^\irred\,,\nn\\
\coeffB{}^\red &=& \coeffB{} - \coeffB{}^\irred\,.\nn
\end{eqnarray}
At the first stage, we solve the homogeneous set
of irreducible equations,
\begin{equation}
\coeffB{}^\irred E^\irred = 0\,.
\label{IrreducibleEquation}
\end{equation}
The full equation can then be rewritten as follows,
\begin{equation}
\coeffB{}^\red E
= -\coeffB{}^\irred E^\red \,,
\label{InhomogeneousReducibleEquation}
\end{equation}
which is an inhomogeneous equation for $\coeffB{}^\red$ in terms
of the (now-known) irreducible polynomials $\coeffB{}^\irred$.  It
turns out to be better to solve these equations in two stages,
first only the rows of $E$ arising from propagators involving
only $\ell_1$ or $\ell_2$, but not both; and adding in the
full set of equations in a second stage.

In order to solve the inhomogeneous equations, we write out an auxiliary
set of equations which impose reducibility on each of the 
coefficients $\coeffB{}^\red$,
\begin{equation}
\coeffB{\alpha} = \sum_{j=1}^{n_d} \coeffB{\alpha,j} 
                 (\sig_{j1}\ell_1+\sig_{j2}\ell_2-K_j)^2\,.
\label{ReducibilityConstraint}
\end{equation}
The sum could also be taken over the Gr\"obner basis elements used
to define the $\Irred$ operator.  There are advantages and disadvantages
to simply adding these equations as auxiliary equations, with additional
unknowns $\coeffB{\alpha,j}$, as opposed to substituting these expansions
into the inhomogeneous equations~(\ref{InhomogeneousReducibleEquation}), and
both the memory usage and the time required for solving can depend
sensitively on this choice.  In the examples we have considered, it
appears better not to substitute in the first of the two stages
of solving the inhomogeneous equations, and better to substitute in
the second stage.

At the first stage, we select the columns 
in \eqn{InhomogeneousReducibleEquation} corresponding to propagators
involving either $\ell_1$ or $\ell_2$, but not both; in the planar
case, this means all but the last column.  Call the solutions
of these equations $\coeffB{\alpha}^{\rm II}$.  At the second stage, 
we split the $\coeffB{\alpha}$,
\begin{equation}
\coeffB{\alpha} = \coeffB{\alpha}^{\rm II} + \delta \coeffB{\alpha}^\red\,,
\label{SplitSolution}
\end{equation}
and solve \eqn{InhomogeneousReducibleEquation} for
$\delta\coeffB{}^\red$, with all but the last column of the right-hand
side replaced by zero, and with an additional reducibility constraint
of the form~(\ref{ReducibilityConstraint}) imposed on $\delta\coeffB{}^\red$.

The strategy for solving the inhomogeneous 
equations~(\ref{InhomogeneousEquation}) can
be summarized in the following algorithm (Algorithm II),
\begin{itemize}
\item[1.] Compute the Gr\"obner basis $G$ of the rows of $E$ (it is 
 sufficient to compute a partial Gr\"obner basis limited to the maximal
 engineering dimension found in the right-hand side $f$), along with
 the cofactor matrix $C$;

\item[2.] Reduce the right-hand side $f$ over $G$, yield coefficients
$q_f$ and a remainder $r_f$;

\item[3.] If the remainder $r_f$ of this reduction does not vanish identically,
solve the equation $r_f-\tilde g G = 0$ with dummy coefficients $\tilde g$
using Algorithm I (in practice, it is best to impose an engineering dimension 
limit in the intermediate steps of Algorithm I, and increment it until the
solution converges to a stable one);

\item[4.] If a dimension-limited Gr\"obner basis was computed in step 1,
and $r_f$ was not identically zero at step 3, 
extend the Gr\"obner basis to the new (typically larger) 
maximal engineering dimension
in the constrained form of $f$.  Rather than starting from scratch, one
can start from the original Gr\"obner basis $G$, in which case the full
cofactor matrix will be given by the product of the new and old matrices,
$C_f = C_2 C_1$.

\item[5.] The solution is then $q_f C$.
\end{itemize}

The improved strategy for solving the original 
equation~(\ref{IBPMatrixEquation})
can then be summarized in the following algorithm
(Algorithm III),
\begin{itemize}
\item[1.] Compute solutions to \eqn{IrreducibleEquation} using Algorithm~I;

\item[2.] Solve the rows in the 
inhomogeneous equation~(\ref{InhomogeneousReducibleEquation}) 
corresponding to propagators containing a lone loop momentum, for the 
reducible terms $\coeffB{}^\red$, along with the constraint 
equations~(\ref{ReducibilityConstraint}) expressing reducibility,
using Algorithm~II;

\item[3.] Write each coefficient $\coeffB{}^\red$ as a sum of this
solution, and another coefficient $\delta\coeffB{}^\red$,
as in \eqn{SplitSolution}.  Solve
the inhomogeneous equation corresponding to the propagator containing
both loop momenta, along with constraint equations of step 2 with their
right-hand sides set to zero, using Algorithm~II (here it is better
to substitute the reducibility constraint 
equations~(\ref{ReducibilityConstraint}) back into
the inhomogeneous equations);

\item[4.] Reduce to a set of independent solutions with respect
to reduction by the set of propagator denominators.
\end{itemize}

\section{The Massless Double Box}
\label{MasslessDoubleBoxSection}

\begin{figure}[ht]
\includegraphics[scale=0.7]{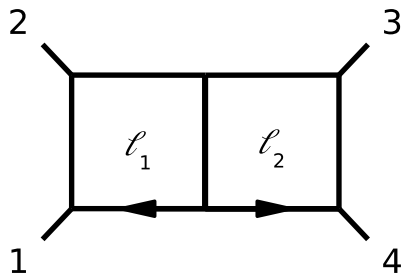}
 \caption{The double box $P^{**}_{2,2}$.} 
\label{MasslessDoubleBoxFigure}
\end{figure}
As an example of how to apply the ideas presented in the previous
section, let us examine the planar double box, $P_{2,2}^{**}$, with all
external legs taken to be massless.  The integral is 
shown in fig.~\ref{MasslessDoubleBoxFigure}.
The
$D$-dimensional reductions were worked out several years ago for the
integral with all external masses vanishing, using \AIR~\cite{AIR}.
The same reductions have been worked out (though not reported explicitly)
for configurations with one  external mass~\cite{GehrmannRemiddi}.
 The three- and four-mass cases
have not been worked out previously.  We discuss the massive cases
in the next section.

As described in the previous section, we
start by looking for vectors $v_{1,2}^\mu$ that give rise to
IBP equations free of doubled propagators.  
For the double box, we choose $k_{1,2,4}$ as basis
momenta, so the general form~(\ref{IBPVectorGeneralForm}) becomes,
\begin{equation}
v_i^\mu = \coeffA{\ell_1}_i \ell_1^\mu
+\coeffA{\ell_2}_i \ell_2^\mu
+\coeffA{1}_i k_1^\mu
+\coeffA{2}_i k_2^\mu
+\coeffA{4}_i k_4^\mu\,,
\label{DoubleBoxGeneralForm}
\end{equation}
where each of the coefficients $\coeffA{p}$ is itself a function
of Lorentz invariants in $V$, which here we can take to be the following
set,
\begin{equation}
V_{22} = \{
\ell_1^2\,, \ell_1\cdot \ell_2, \ell_2^2\,,
\ell_1\cdot k_1\,,
\ell_1\cdot k_2\,,
\ell_1\cdot k_4\,,
\ell_2\cdot k_1\,,
\ell_2\cdot k_3\,,
\ell_2\cdot k_4\,,s_{12}\}\,.
\label{BasicVariablesDoubleBox}
\end{equation}
The other dot products of the loop momenta 
can be expressed in terms of these via
momentum conservation, and the other independent invariant
is treated as a multiple of $s_{12}$, $s_{14} = \chi_{14} s_{12}$.
There are two irreducible numerators,
$\ell_1\cdot k_4$ and $\ell_2\cdot k_1$.

\def\cdott{\!\cdot\!}
\catcode`@=11
\def\tmatrix#1{\null\,\vcenter{
\def\tm{\!-\!}\def\tp{\!+\!}
\lineskip 4pt  \baselineskip 12pt
\m@th
    \ialign{\hfil$##$\hfil&&\quad\hfil$##$\hfil\crcr
      \crcr\noalign{\kern-\baselineskip}
      #1\crcr \crcr\noalign{\kern-\baselineskip}}}\,}
\catcode`@=12 
\def\lfrac#1#2{#1/#2}
The matrix $E$ of \eqn{EDefinition} then takes the form,
\begin{eqnarray}
&&\mathstrut\nn\\
E &=&
8\left(\tmatrix{
\ell_{1}^2\!&\!- k_{1}\!\cdot\!\ell_{1}+\ell_{1}^2\!&\!- k_{1}\!\cdot\!\ell_{1}- k_{2}\!\cdot\!\ell_{1}+\ell_{1}^2\!&\!0\cr
\ell_{1}\!\cdot\!\ell_{2}\!&\!- k_{1}\!\cdot\!\ell_{2}+\ell_{1}\!\cdot\!\ell_{2}\!&\!k_{3}\!\cdot\!\ell_{2}+k_{4}\!\cdot\!\ell_{2}+\ell_{1}\!\cdot\!\ell_{2}\!&\!0\cr
k_{1}\!\cdot\!\ell_{1}\!&\!k_{1}\!\cdot\!\ell_{1}\!&\!k_{1}\!\cdot\!\ell_{1}-\lfrac{s_{12}}{2}\!&\!0\cr
k_{2}\!\cdot\!\ell_{1}\!&\!k_{2}\!\cdot\!\ell_{1}-\lfrac{s_{12}}{2}\!&\!k_{2}\!\cdot\!\ell_{1}-\lfrac{s_{12}}{2}\!&\!0\cr
k_{4}\!\cdot\!\ell_{1}\!&\!k_{4}\!\cdot\!\ell_{1}-\lfrac{\chi_{14}s_{12}}{2}\!&\!k_{4}\!\cdot\!\ell_{1}+\lfrac{s_{12}}{2}\!&\!0\cr
0\!&\!0\!&\!0\!&\!\ell_{1}\!\cdot\!\ell_{2}\cr
0\!&\!0\!&\!0\!&\!\ell_{2}^2\cr
0\!&\!0\!&\!0\!&\!k_{1}\!\cdot\!\ell_{2}\cr
0\!&\!0\!&\!0\!&\!\hskip -4mm - k_{1}\!\cdot\!\ell_{2}- k_{3}\!\cdot\!\ell_{2}- k_{4}\!\cdot\!\ell_{2}\cr
0\!&\!0\!&\!0\!&\!k_{4}\!\cdot\!\ell_{2}\cr
\lfrac{\ell_{1}^2}{4}\!&\!0\!&\!0\!&\!0\cr
0\!&\!\lfrac{\ell_{1}^2}{4}-\lfrac{k_{1}\!\cdot\!\ell_{1}}{2}\!&\!0\!&\!0\cr
0\!&\!0\!&\!\hskip -4mm\lfrac{\ell_{1}^2}{4}+\lfrac{s_{12}}{4}-\lfrac{k_{1}\!\cdot\!\ell_{1}}{2}-\lfrac{k_{2}\!\cdot\!\ell_{1}}{2}\!&\!0\cr
0\!&\!0\!&\!0\!&\!\lfrac{\ell_{2}^2}{4}\cr
0\!&\!0\!&\!0\!&\!0\cr
0\!&\!0\!&\!0\!&\!0\cr
0\!&\!0\!&\!0\!&\!0
\cr
}\right.\,\nn\\
&&\mathstrut\\
&&\mathstrut\nn\\
&& \hskip -8mm \left.\tmatrix{
0\!&\!0\!&\!\ell_{1}^2+\ell_{1}\!\cdot\!\ell_{2}\cr
0\!&\!0\!&\!\ell_{1}\!\cdot\!\ell_{2}+\ell_{2}^2\cr
0\!&\!0\!&\!k_{1}\!\cdot\!\ell_{1}+k_{1}\!\cdot\!\ell_{2}\cr
0\!&\!0\!&\!k_{2}\!\cdot\!\ell_{1}\tm k_{1}\!\cdot\!\ell_{2}\tm k_{3}\!\cdot\!\ell_{2}\tm k_{4}\!\cdot\!\ell_{2}\cr
0\!&\!0\!&\!k_{4}\!\cdot\!\ell_{1}+k_{4}\!\cdot\!\ell_{2}\cr
- k_{4}\!\cdot\!\ell_{1}+\ell_{1}\!\cdot\!\ell_{2}\!&\!k_{1}\!\cdot\!\ell_{1}+k_{2}\!\cdot\!\ell_{1}+\ell_{1}\!\cdot\!\ell_{2}\!&\!\ell_{1}^2+\ell_{1}\!\cdot\!\ell_{2}\cr
- k_{4}\!\cdot\!\ell_{2}+\ell_{2}^2\!&\!- k_{3}\!\cdot\!\ell_{2}- k_{4}\!\cdot\!\ell_{2}+\ell_{2}^2\!&\!\ell_{1}\!\cdot\!\ell_{2}+\ell_{2}^2\cr
k_{1}\!\cdot\!\ell_{2}-\lfrac{\chi_{14}s_{12}}{2}\!&\!k_{1}\!\cdot\!\ell_{2}+\lfrac{s_{12}}{2}\!&\!k_{1}\!\cdot\!\ell_{1}+k_{1}\!\cdot\!\ell_{2}\cr
\lfrac{(1\tp\chi_{14})s_{12}}{2}\tm k_{1}\!\cdot\!\ell_{2}\tm k_{3}\!\cdot\!\ell_{2}\tm k_{4}\!\cdot\!\ell_{2}\!&\!\lfrac{s_{12}}{2}\tm k_{1}\!\cdot\!\ell_{2}\tm k_{3}\!\cdot\!\ell_{2}\tm k_{4}\!\cdot\!\ell_{2}\!&\!k_{2}\!\cdot\!\ell_{1}\tm k_{1}\!\cdot\!\ell_{2}\tm k_{3}\!\cdot\!\ell_{2}\tm k_{4}\!\cdot\!\ell_{2}\cr
k_{4}\!\cdot\!\ell_{2}\!&\!k_{4}\!\cdot\!\ell_{2}-\lfrac{s_{12}}{2}\!&\!k_{4}\!\cdot\!\ell_{1}+k_{4}\!\cdot\!\ell_{2}\cr
0\!&\!0\!&\!0\cr
0\!&\!0\!&\!0\cr
0\!&\!0\!&\!0\cr
0\!&\!0\!&\!0\cr
-\lfrac{k_{4}\!\cdot\!\ell_{2}}{2}+\lfrac{\ell_{2}^2}{4}\!&\!0\!&\!0\cr
0\!&\!\hskip -6mm\lfrac{\ell_{2}^2}{4}\tp\lfrac{s_{12}}{4}\tm\lfrac{k_{3}\!\cdot\!\ell_{2}}{2}\tm\lfrac{k_{4}\!\cdot\!\ell_{2}}{2}\!&\!0\cr
0\!&\!0\!&\!\lfrac{\ell_{1}^2}{4}+\lfrac{\ell_{1}\!\cdot\!\ell_{2}}{2}+\lfrac{\ell_{2}^2}{4}
\cr}\right)\,,\nn\\
&&\mathstrut\nn
\end{eqnarray}

while the vector $\coeffB{}$ is,
\begin{equation}
\coeffB{} = 
\bigl(
\coeffA{\ell_1}_1\,\coeffA{\ell_2}_1\,\coeffA{1}_1\,\coeffA{2}_1\,\coeffA{4}_1\,\coeffA{\ell_1}_2\,\coeffA{\ell_2}_2\,\coeffA{1}_2\,\coeffA{2}_2\,\coeffA{4}_2\,
u_{1\ldots7}\bigr)\,.
\end{equation}
The leftmost three columns correspond to equations for the left-hand loop;
the following three columns, equations for the right-hand loop; and
the last column, an equation for the common propagator.

The propagator denominators are,
\begin{equation}
\ell_1^2\,;\,\,
(\ell_1-k_1)^2\,;\,\,
(\ell_1-K_{12})^2\,;\,\,
\ell_2^2\,;\,\,
(\ell_2-k_4)^2\,;\,\,
(\ell_2-K_{34})^2\,;\,\,
(\ell_1+\ell_2)^2\,.
\end{equation}
Reducibility of an expression may be determined by reducing
over a Gr\"obner basis of these denominators, which
using a plain lexical ordering is,
\begin{equation}
\ell_1\tcdot\ell_2\,,
\ell_2^2\,,
k_4\tcdot\ell_2\,,
k_3\tcdot\ell_2-s_{12}/2\,,
\ell_1^2\,,
k_2\tcdot\ell_1-s_{12}/2\,,
k_1\tcdot\ell_1\,.
\label{PropagatorGroebnerBasis}
\end{equation}
The $\Irred$ operator gives the remainder after this reduction; for
example,
\begin{eqnarray}
&&\Irred \bigl(
a_1 \ell_1^2 + a_2 \ell_1\tcdot \ell_2
+a_3 \ell_2^2+a_4 \ell_1\tcdot k_1+a_5 \ell_1\tcdot k_2+a_6 \ell_1\tcdot k_4
+a_7 \ell_2\tcdot k_1+a_8 \ell_2\tcdot k_3\nn\\
&&\hskip 10mm+a_9 \ell_2\tcdot k_4
+a_{10} s_{12}/2
\bigr)
\nn\\
&&= 
(a_5+a_8+a_{10}) s_{12}/2 + a_6 k_4\tcdot\ell_1 + a_7 k_1\tcdot\ell_2\,.
\end{eqnarray}

When using algorithm~I from the previous section to solve 
\eqn{IBPMatrixEquation}, we first obtain a Gr\"obner basis
for the seven-tuples making up the rows of $E$.  There are
50 tuples in this basis, which in turn give rise to 
167 syzygies, which we can represent as 17-tuples.  Polynomial
reduction and removal of completely-reducible syzygies (with respect
to the basis in \eqn{PropagatorGroebnerBasis}) leaves us
with 119 syzygies of the Gr\"obner basis.  These in turn give
rise to 101 solutions of \eqn{IBPMatrixEquation}, 
of which three are independent.  The matrix
$I-Q C$ gives no additional solutions (which can be understood on
dimensional grounds here).

There is one solution whose coefficients are of engineering dimension
two,
\begin{eqnarray}
v_{1;1} &=&
 -2 ( k_{4}\tcdot \ell_{1}+\ell_{1}^2)k_{1}^\mu- \ell_{1}^2 k_{2}^\mu
 +(2 k_{1}\tcdot \ell_{1}- \ell_{1}^2)k_{4}^\mu+(4 k_{1}\tcdot \ell_{1}
+2 k_{2}\tcdot \ell_{1}+2 k_{4}\tcdot \ell_{1}- s_{12})\ell_{1}^\mu\,,\nn\\
v_{1;2} &=&
2 (\ell_{2}^2- k_{4}\tcdot \ell_{2})k_{1}^\mu+\ell_{2}^2k_{2}^\mu
+(2 k_{1}\tcdot \ell_{2}+\ell_{2}^2)k_{4}^\mu
+(2 k_{3}\tcdot \ell_{2}-2 k_{1}\tcdot \ell_{2}- s_{12})\ell_{2}^\mu\,;
\label{MasslessDoubleBoxVector1}
\end{eqnarray}
and two solutions with coefficients of engineering dimension four,
\begin{eqnarray}
v_{2;1} &=&
(-4 k_{2}\!\cdot\!\ell_{1}k_{4}\!\cdot\!\ell_{1}-4 k_{3}\!\cdot\!\ell_{2}\ell_{1}^2+4 k_{4}\!\cdot\!\ell_{1}\ell_{1}^2-4 k_{4}\!\cdot\!\ell_{2}\ell_{1}^2-4 \ell_{1}^2\ell_{1}\!\cdot\!\ell_{2}-2 \ell_{1}^2\ell_{2}^2-2 \chi_{14}\ell_{1}^2s_{12})k_{1}^\mu\nn\\
&&+(4 k_{1}\!\cdot\!\ell_{1}k_{4}\!\cdot\!\ell_{1}-2 k_{1}\!\cdot\!\ell_{1}\ell_{1}^2-2 k_{2}\!\cdot\!\ell_{1}\ell_{1}^2-4 k_{3}\!\cdot\!\ell_{2}\ell_{1}^2-4 k_{4}\!\cdot\!\ell_{2}\ell_{1}^2-4 \ell_{1}^2\ell_{1}\!\cdot\!\ell_{2}-2 \ell_{1}^2\ell_{2}^2\nn\\
&&+2 \ell_{1}^2s_{12}-2 \chi_{14}\ell_{1}^2s_{12})k_{2}^\mu+(-4 k_{1}\!\cdot\!\ell_{1}\ell_{1}^2-4 k_{2}\!\cdot\!\ell_{1}\ell_{1}^2+2 (\ell_{1}^2)^{2}+2 \ell_{1}^2s_{12})k_{4}^\mu\nn\\
&&+(4 k_{1}\!\cdot\!\ell_{1}k_{2}\!\cdot\!\ell_{1}+4 (k_{2}\!\cdot\!\ell_{1})^{2}+8 k_{1}\!\cdot\!\ell_{1}k_{3}\!\cdot\!\ell_{2}+8 k_{2}\!\cdot\!\ell_{1}k_{3}\!\cdot\!\ell_{2}+8 k_{2}\!\cdot\!\ell_{1}k_{4}\!\cdot\!\ell_{1}+8 k_{1}\!\cdot\!\ell_{1}k_{4}\!\cdot\!\ell_{2}\nn\\
&&+8 k_{2}\!\cdot\!\ell_{1}k_{4}\!\cdot\!\ell_{2}-4 k_{4}\!\cdot\!\ell_{1}\ell_{1}^2+8 k_{1}\!\cdot\!\ell_{1}\ell_{1}\!\cdot\!\ell_{2}+8 k_{2}\!\cdot\!\ell_{1}\ell_{1}\!\cdot\!\ell_{2}+4 k_{1}\!\cdot\!\ell_{1}\ell_{2}^2+4 k_{2}\!\cdot\!\ell_{1}\ell_{2}^2\nn\\
&&-4 k_{1}\!\cdot\!\ell_{1}s_{12}-6 k_{2}\!\cdot\!\ell_{1}s_{12}-4 k_{3}\!\cdot\!\ell_{2}s_{12}-2 k_{4}\!\cdot\!\ell_{1}s_{12}-4 k_{4}\!\cdot\!\ell_{2}s_{12}+\ell_{1}^2s_{12}+2 \chi_{14}\ell_{1}^2s_{12}\nn\\
&&-4 \ell_{1}\!\cdot\!\ell_{2}s_{12}-2 \ell_{2}^2s_{12}+2 s_{12}^{2})\ell_{1}^\mu\,,
\label{MasslessDoubleBoxVector2a}\\
v_{2;2} &=&
(4 k_{1}\!\cdot\!\ell_{2}k_{4}\!\cdot\!\ell_{1}+4 k_{3}\!\cdot\!\ell_{2}k_{4}\!\cdot\!\ell_{1}+4 k_{4}\!\cdot\!\ell_{1}k_{4}\!\cdot\!\ell_{2}-4 k_{4}\!\cdot\!\ell_{2}\ell_{1}\!\cdot\!\ell_{2}+4 k_{3}\!\cdot\!\ell_{2}\ell_{2}^2-4 k_{4}\!\cdot\!\ell_{1}\ell_{2}^2\nn\\
&&+6 \ell_{1}\!\cdot\!\ell_{2}\ell_{2}^2+4 (\ell_{2}^2)^{2}-2 \ell_{1}\!\cdot\!\ell_{2}s_{12}-2 \chi_{14}\ell_{1}\!\cdot\!\ell_{2}s_{12}-2 \ell_{2}^2s_{12})k_{1}^\mu+(4 k_{1}\!\cdot\!\ell_{2}k_{4}\!\cdot\!\ell_{1}\nn\\
&&-4 k_{4}\!\cdot\!\ell_{2}\ell_{1}\!\cdot\!\ell_{2}+2 k_{1}\!\cdot\!\ell_{1}\ell_{2}^2+2 k_{2}\!\cdot\!\ell_{1}\ell_{2}^2+4 k_{3}\!\cdot\!\ell_{2}\ell_{2}^2+6 \ell_{1}\!\cdot\!\ell_{2}\ell_{2}^2+4 (\ell_{2}^2)^{2}-2 \chi_{14}\ell_{1}\!\cdot\!\ell_{2}s_{12}\nn\\
&&-2 \ell_{2}^2s_{12})k_{2}^\mu+(4 k_{1}\!\cdot\!\ell_{1}\ell_{2}^2+4 k_{2}\!\cdot\!\ell_{1}\ell_{2}^2+4 \ell_{1}\!\cdot\!\ell_{2}\ell_{2}^2+2 (\ell_{2}^2)^{2}-2 \ell_{2}^2s_{12})k_{4}^\mu\nn\\
&&+(-4 k_{3}\!\cdot\!\ell_{2}k_{4}\!\cdot\!\ell_{2}-4 (k_{4}\!\cdot\!\ell_{2})^{2}+2 k_{3}\!\cdot\!\ell_{2}\ell_{2}^2+2 k_{4}\!\cdot\!\ell_{2}\ell_{2}^2+2 k_{1}\!\cdot\!\ell_{2}s_{12}-2 \chi_{14}k_{3}\!\cdot\!\ell_{2}s_{12}\nn\\
&&-2 \chi_{14}k_{4}\!\cdot\!\ell_{2}s_{12}+\ell_{2}^2s_{12}+2 \chi_{14}\ell_{2}^2s_{12})\ell_{1}^\mu+(4 k_{1}\!\cdot\!\ell_{1}k_{1}\!\cdot\!\ell_{2}+4 k_{1}\!\cdot\!\ell_{2}k_{2}\!\cdot\!\ell_{1}+4 k_{1}\!\cdot\!\ell_{1}k_{3}\!\cdot\!\ell_{2}\nn\\
&&+4 k_{2}\!\cdot\!\ell_{1}k_{3}\!\cdot\!\ell_{2}+8 (k_{3}\!\cdot\!\ell_{2})^{2}+8 k_{1}\!\cdot\!\ell_{2}k_{4}\!\cdot\!\ell_{1}+8 k_{3}\!\cdot\!\ell_{2}k_{4}\!\cdot\!\ell_{2}+8 k_{3}\!\cdot\!\ell_{2}\ell_{1}\!\cdot\!\ell_{2}-2 k_{1}\!\cdot\!\ell_{1}\ell_{2}^2\nn\\
&&-2 k_{2}\!\cdot\!\ell_{1}\ell_{2}^2+8 k_{3}\!\cdot\!\ell_{2}\ell_{2}^2+4 k_{4}\!\cdot\!\ell_{2}\ell_{2}^2-2 k_{1}\!\cdot\!\ell_{1}s_{12}-2 \chi_{14}k_{1}\!\cdot\!\ell_{1}s_{12}-2 k_{2}\!\cdot\!\ell_{1}s_{12}\nn\\
&&-2 \chi_{14}k_{2}\!\cdot\!\ell_{1}s_{12}-8 k_{3}\!\cdot\!\ell_{2}s_{12}+2 k_{4}\!\cdot\!\ell_{1}s_{12}-4 k_{4}\!\cdot\!\ell_{2}s_{12}-6 \ell_{1}\!\cdot\!\ell_{2}s_{12}-4 \chi_{14}\ell_{1}\!\cdot\!\ell_{2}s_{12}\nn\\
&&-2 \ell_{2}^2s_{12}+2 s_{12}^{2})\ell_{2}^\mu\,;\nn
%
\end{eqnarray}
and
\begin{eqnarray}
v_{3;1} &=&
(-4 k_{1}\!\cdot\!\ell_{2}\ell_{1}^2-\frac{2 k_{1}\!\cdot\!\ell_{2}\ell_{1}^2}{\chi_{14}}-2 (\ell_{1}^2)^{2}-\frac{ (\ell_{1}^2)^{2}}{\chi_{14}}-4 \ell_{1}^2\ell_{1}\!\cdot\!\ell_{2}-\frac{2 \ell_{1}^2\ell_{1}\!\cdot\!\ell_{2}}{\chi_{14}}-2 k_{2}\!\cdot\!\ell_{1}\ell_{2}^2-\frac{ \ell_{1}^2\ell_{2}^2}{\chi_{14}}\nn\\
&&+\chi_{14}\ell_{1}^2s_{12}-2 \ell_{1}\!\cdot\!\ell_{2}s_{12}-2 \chi_{14}\ell_{1}\!\cdot\!\ell_{2}s_{12})k_{1}^\mu+(4 k_{1}\!\cdot\!\ell_{2}k_{4}\!\cdot\!\ell_{1}-2 k_{1}\!\cdot\!\ell_{2}\ell_{1}^2-\frac{2 k_{1}\!\cdot\!\ell_{2}\ell_{1}^2}{\chi_{14}}\nn\\
&&+2 k_{3}\!\cdot\!\ell_{2}\ell_{1}^2+2 k_{4}\!\cdot\!\ell_{2}\ell_{1}^2-2 (\ell_{1}^2)^{2}-\frac{ (\ell_{1}^2)^{2}}{\chi_{14}}-4 \ell_{1}^2\ell_{1}\!\cdot\!\ell_{2}-\frac{2 \ell_{1}^2\ell_{1}\!\cdot\!\ell_{2}}{\chi_{14}}+2 k_{1}\!\cdot\!\ell_{1}\ell_{2}^2-2 \ell_{1}^2\ell_{2}^2-\frac{ \ell_{1}^2\ell_{2}^2}{\chi_{14}}\nn\\
&&+\chi_{14}\ell_{1}^2s_{12}-2 \chi_{14}\ell_{1}\!\cdot\!\ell_{2}s_{12})k_{2}^\mu+(-4 k_{1}\!\cdot\!\ell_{2}k_{2}\!\cdot\!\ell_{1}+2 k_{1}\!\cdot\!\ell_{2}\ell_{1}^2+2 k_{3}\!\cdot\!\ell_{2}\ell_{1}^2+2 k_{4}\!\cdot\!\ell_{2}\ell_{1}^2\nn\\
&&+2 \ell_{1}\!\cdot\!\ell_{2}s_{12})k_{4}^\mu+(8 k_{1}\!\cdot\!\ell_{1}k_{1}\!\cdot\!\ell_{2}+\frac{4 k_{1}\!\cdot\!\ell_{1}k_{1}\!\cdot\!\ell_{2}}{\chi_{14}}+4 k_{1}\!\cdot\!\ell_{2}k_{2}\!\cdot\!\ell_{1}+\frac{4 k_{1}\!\cdot\!\ell_{2}k_{2}\!\cdot\!\ell_{1}}{\chi_{14}}\nn\\
&&-4 k_{2}\!\cdot\!\ell_{1}k_{3}\!\cdot\!\ell_{2}-4 k_{1}\!\cdot\!\ell_{2}k_{4}\!\cdot\!\ell_{1}-4 k_{3}\!\cdot\!\ell_{2}k_{4}\!\cdot\!\ell_{1}-4 k_{2}\!\cdot\!\ell_{1}k_{4}\!\cdot\!\ell_{2}-4 k_{4}\!\cdot\!\ell_{1}k_{4}\!\cdot\!\ell_{2}+4 k_{1}\!\cdot\!\ell_{1}\ell_{1}^2\nn\\
&&+\frac{2 k_{1}\!\cdot\!\ell_{1}\ell_{1}^2}{\chi_{14}}+4 k_{2}\!\cdot\!\ell_{1}\ell_{1}^2+\frac{2 k_{2}\!\cdot\!\ell_{1}\ell_{1}^2}{\chi_{14}}+8 k_{1}\!\cdot\!\ell_{1}\ell_{1}\!\cdot\!\ell_{2}+\frac{4 k_{1}\!\cdot\!\ell_{1}\ell_{1}\!\cdot\!\ell_{2}}{\chi_{14}}+8 k_{2}\!\cdot\!\ell_{1}\ell_{1}\!\cdot\!\ell_{2}\nn\\
&&+\frac{4 k_{2}\!\cdot\!\ell_{1}\ell_{1}\!\cdot\!\ell_{2}}{\chi_{14}}+\frac{2 k_{1}\!\cdot\!\ell_{1}\ell_{2}^2}{\chi_{14}}+4 k_{2}\!\cdot\!\ell_{1}\ell_{2}^2+\frac{2 k_{2}\!\cdot\!\ell_{1}\ell_{2}^2}{\chi_{14}}-2 \chi_{14}k_{1}\!\cdot\!\ell_{1}s_{12}-4 k_{1}\!\cdot\!\ell_{2}s_{12}\nn\\
&&-\frac{2 k_{1}\!\cdot\!\ell_{2}s_{12}}{\chi_{14}}-2 \chi_{14}k_{2}\!\cdot\!\ell_{1}s_{12}+2 k_{3}\!\cdot\!\ell_{2}s_{12}+2 \chi_{14}k_{3}\!\cdot\!\ell_{2}s_{12}+2 k_{4}\!\cdot\!\ell_{2}s_{12}+2 \chi_{14}k_{4}\!\cdot\!\ell_{2}s_{12}\nn\\
&&-2 \ell_{1}^2s_{12}-\frac{ \ell_{1}^2s_{12}}{\chi_{14}}-\frac{2 \ell_{1}\!\cdot\!\ell_{2}s_{12}}{\chi_{14}}+4 \chi_{14}\ell_{1}\!\cdot\!\ell_{2}s_{12}- \ell_{2}^2s_{12}-\frac{ \ell_{2}^2s_{12}}{\chi_{14}}+\chi_{14}s_{12}^{2})\ell_{1}^\mu+(2 k_{1}\!\cdot\!\ell_{1}s_{12}\nn\\
&&+2 \chi_{14}k_{1}\!\cdot\!\ell_{1}s_{12}+2 \chi_{14}k_{2}\!\cdot\!\ell_{1}s_{12}-2 k_{4}\!\cdot\!\ell_{1}s_{12}-2 \ell_{1}^2s_{12}-2 \chi_{14}\ell_{1}^2s_{12})\ell_{2}^\mu\,,
\label{MasslessDoubleBoxVector2b}\\
v_{3;2} &=&
(-4 k_{4}\!\cdot\!\ell_{2}\ell_{1}^2-\frac{2 k_{4}\!\cdot\!\ell_{2}\ell_{1}^2}{\chi_{14}}-8 k_{4}\!\cdot\!\ell_{2}\ell_{1}\!\cdot\!\ell_{2}-\frac{4 k_{4}\!\cdot\!\ell_{2}\ell_{1}\!\cdot\!\ell_{2}}{\chi_{14}}+6 k_{1}\!\cdot\!\ell_{2}\ell_{2}^2+\frac{2 k_{1}\!\cdot\!\ell_{2}\ell_{2}^2}{\chi_{14}}+2 k_{3}\!\cdot\!\ell_{2}\ell_{2}^2\nn\\
&&-2 k_{4}\!\cdot\!\ell_{2}\ell_{2}^2-\frac{2 k_{4}\!\cdot\!\ell_{2}\ell_{2}^2}{\chi_{14}}+4 \ell_{1}^2\ell_{2}^2+\frac{2 \ell_{1}^2\ell_{2}^2}{\chi_{14}}+8 \ell_{1}\!\cdot\!\ell_{2}\ell_{2}^2+\frac{4 \ell_{1}\!\cdot\!\ell_{2}\ell_{2}^2}{\chi_{14}}+2 (\ell_{2}^2)^{2}+\frac{2 (\ell_{2}^2)^{2}}{\chi_{14}}\nn\\
&&-2 \ell_{2}^2s_{12}-3 \chi_{14}\ell_{2}^2s_{12})k_{1}^\mu+(4 k_{1}\!\cdot\!\ell_{2}k_{4}\!\cdot\!\ell_{2}+4 k_{1}\!\cdot\!\ell_{2}\ell_{2}^2+\frac{2 k_{1}\!\cdot\!\ell_{2}\ell_{2}^2}{\chi_{14}}-2 k_{3}\!\cdot\!\ell_{2}\ell_{2}^2-2 k_{4}\!\cdot\!\ell_{2}\ell_{2}^2\nn\\
&&+2 \ell_{1}^2\ell_{2}^2+\frac{\ell_{1}^2\ell_{2}^2}{\chi_{14}}+4 \ell_{1}\!\cdot\!\ell_{2}\ell_{2}^2+\frac{2 \ell_{1}\!\cdot\!\ell_{2}\ell_{2}^2}{\chi_{14}}+2 (\ell_{2}^2)^{2}+\frac{(\ell_{2}^2)^{2}}{\chi_{14}}-3 \chi_{14}\ell_{2}^2s_{12})k_{2}^\mu+(4 (k_{1}\!\cdot\!\ell_{2})^{2}\nn\\
&&+4 k_{1}\!\cdot\!\ell_{2}k_{3}\!\cdot\!\ell_{2}+4 k_{1}\!\cdot\!\ell_{2}k_{4}\!\cdot\!\ell_{2}+4 k_{1}\!\cdot\!\ell_{2}\ell_{1}^2+\frac{2 k_{1}\!\cdot\!\ell_{2}\ell_{1}^2}{\chi_{14}}+8 k_{1}\!\cdot\!\ell_{2}\ell_{1}\!\cdot\!\ell_{2}+\frac{4 k_{1}\!\cdot\!\ell_{2}\ell_{1}\!\cdot\!\ell_{2}}{\chi_{14}}\nn\\
&&+2 k_{1}\!\cdot\!\ell_{2}\ell_{2}^2+\frac{2 k_{1}\!\cdot\!\ell_{2}\ell_{2}^2}{\chi_{14}}-2 k_{3}\!\cdot\!\ell_{2}\ell_{2}^2-2 k_{4}\!\cdot\!\ell_{2}\ell_{2}^2+2 \ell_{1}^2\ell_{2}^2+\frac{\ell_{1}^2\ell_{2}^2}{\chi_{14}}+4 \ell_{1}\!\cdot\!\ell_{2}\ell_{2}^2+\frac{2 \ell_{1}\!\cdot\!\ell_{2}\ell_{2}^2}{\chi_{14}}\nn\\
&&+2 (\ell_{2}^2)^{2}+\frac{(\ell_{2}^2)^{2}}{\chi_{14}}+2 \ell_{2}^2s_{12})k_{4}^\mu+(-4 (k_{1}\!\cdot\!\ell_{2})^{2}+\frac{4 k_{1}\!\cdot\!\ell_{2}k_{3}\!\cdot\!\ell_{2}}{\chi_{14}}-4 (k_{3}\!\cdot\!\ell_{2})^{2}+4 k_{1}\!\cdot\!\ell_{2}k_{4}\!\cdot\!\ell_{2}\nn\\
&&+\frac{4 k_{1}\!\cdot\!\ell_{2}k_{4}\!\cdot\!\ell_{2}}{\chi_{14}}-4 k_{3}\!\cdot\!\ell_{2}k_{4}\!\cdot\!\ell_{2}-4 k_{1}\!\cdot\!\ell_{2}\ell_{1}^2-\frac{2 k_{1}\!\cdot\!\ell_{2}\ell_{1}^2}{\chi_{14}}+4 k_{3}\!\cdot\!\ell_{2}\ell_{1}^2+\frac{2 k_{3}\!\cdot\!\ell_{2}\ell_{1}^2}{\chi_{14}}\nn\\
&&-8 k_{1}\!\cdot\!\ell_{2}\ell_{1}\!\cdot\!\ell_{2}-\frac{4 k_{1}\!\cdot\!\ell_{2}\ell_{1}\!\cdot\!\ell_{2}}{\chi_{14}}+8 k_{3}\!\cdot\!\ell_{2}\ell_{1}\!\cdot\!\ell_{2}+\frac{4 k_{3}\!\cdot\!\ell_{2}\ell_{1}\!\cdot\!\ell_{2}}{\chi_{14}}-\frac{2 k_{1}\!\cdot\!\ell_{2}\ell_{2}^2}{\chi_{14}}+4 k_{3}\!\cdot\!\ell_{2}\ell_{2}^2\nn\\
&&+\frac{2 k_{3}\!\cdot\!\ell_{2}\ell_{2}^2}{\chi_{14}}-2 k_{1}\!\cdot\!\ell_{2}s_{12}-\frac{2 k_{1}\!\cdot\!\ell_{2}s_{12}}{\chi_{14}}+2 k_{3}\!\cdot\!\ell_{2}s_{12}-2 \chi_{14}k_{3}\!\cdot\!\ell_{2}s_{12}-2 \chi_{14}k_{4}\!\cdot\!\ell_{2}s_{12}\nn\\
&&-2 \ell_{1}^2s_{12}-\frac{ \ell_{1}^2s_{12}}{\chi_{14}}-4 \ell_{1}\!\cdot\!\ell_{2}s_{12}-\frac{2 \ell_{1}\!\cdot\!\ell_{2}s_{12}}{\chi_{14}}-3 \ell_{2}^2s_{12}-\frac{ \ell_{2}^2s_{12}}{\chi_{14}}-2 \chi_{14}\ell_{2}^2s_{12}+\chi_{14}s_{12}^{2})\ell_{2}^\mu\,.\nn
\end{eqnarray}
The algorithms described in the previous section are not guaranteed to
yield the solutions in the simplest possible form; it can happen that
linear combinations of solutions can be factored to yield a solution
of lower engineering dimension.  In this case, however, the solutions
do appear to be close to the ``simplest'' possible ones.

Were we to use algorithm~III, we would start by solving the 
equation for the irreducible part of $E$.  Here, the Gr\"obner basis
has 12 vectors, giving rise to five syzygies, and three solutions
to \eqn{IrreducibleEquation} --- one with coefficients
of engineering dimension two, the other two with coefficients
of engineering dimension four.  At the second stage, we have 103
basis tuples in the Gr\"obner basis (limited to engineering dimension six)
for $E$ augmented by the auxiliary
equations~(\ref{ReducibilityConstraint}) imposing reducibility.
We find that that the right-hand side of~\eqn{InhomogeneousReducibleEquation}
can be decomposed over this basis, so that $r_f$ in 
\eqn{ReduceOverGroebner} vanishes.  There are again three solutions
to the equations.  At the third stage, we include the last column
of $E$, corresponding to the propagator involving both loop momenta;
this yields 125 tuples in the Gr\"obner basis, and the right-hand
side again reduces over this basis, and we end up with the three
solutions prefigured by the solutions to the irreducible equations.
In more complex integrals, the third stage will often impose additional
constraints on the free polynomials obtained at the first stage,
leading to fewer solutions, or more solutions with coefficients
of higher engineering dimension.

The two algorithms are not guaranteed to produce the same solutions,
and even for the massless double box, they do not.  The solutions
will however span the same space, and yield the same solutions to the
IBP equations for the integrals of interest.
In this case, they do produce the same number of solutions
of each engineering dimension, and the solutions are equivalent --- the
solutions produced by algorithm~I can be written as linear combinations
of those produced by algorithm~III and vice versa.

With the IBP generating vectors of eqns.~(\ref{MasslessDoubleBoxVector1})
and~(\ref{MasslessDoubleBoxVector2a}), we can now construct IBP equations, 
\begin{equation}
0 = P^{**}_{2,2}\biggl[
{\partial\over \partial\ell_1^\mu} p v_{i;1}^\mu
+{\partial\over \partial\ell_2^\mu} p v_{i;2}^\mu
\biggr]\,,
\end{equation}
where $p$ is an irreducible polynomial in the symbols in $V_{22}$.  These
equations will relate various nominally-irreducible $P^{**}_{2,2}$
to integrals with fewer propagators, but by construction
will involve no undesired integrals.  In practice, we do not need
all three solutions; the first two suffice to produce all possible
IBP equations.

The first solution~(\ref{MasslessDoubleBoxVector1}) leads to the
following equation,
\begin{eqnarray}
0 &=& 
2 P^{**}_{2,2}[k_{1}\!\cdot\!\ell_{2}](k_{1},k_{2},k_{3},k_{4})-2 P^{**}_{2,2}[k_{4}\!\cdot\!\ell_{1}](k_{1},k_{2},k_{3},k_{4})
+P^{*}_{2,1}[1](k_{1},k_{2},k_{3},k_{4})
\label{FirstDoubleBoxIBP}\\
&&
-2 P^{*}_{2,1}[1](k_{3},k_{4},k_{1},k_{2})+P^{*}_{2,1}[1](k_{4},k_{3},k_{2},k_{1})- P^{**}_{2,1}[1](k_{1},k_{2},K_{34})+P^{**}_{2,1}[1](k_{4},k_{3},K_{12})
\nn
\end{eqnarray}
If we make use of the symmetries of the 
reduced integrals $P^*_{2,1}$ and $P^{**}_{2,1}$, or reduce these
latter integrals in turn, this equation simplifies
to,
\begin{equation}
P^{**}_{2,2}[k_{1}\cdot\ell_{2}](k_{1},k_{2},k_{3},k_{4})
 =P^{**}_{2,2}[k_{4}\cdot\ell_{1}](k_{1},k_{2},k_{3},k_{4})\,.
\label{FirstDoubleBoxIBPsimp}
\end{equation}
For the all-massless double box, this equation is also a direct consequence
of the symmetries of the integral; but the analogous statement is no longer
true for double boxes with external masses.

We can also use these vectors to derive equations for double boxes
with more complicated numerator insertions, of powers or products of
the basic irreducible numerators.  As discussed earlier,
we can do so by multiplying the vector
by powers of invariants, which still yields a solution to
the equations requiring that the IBP be free of doubled propagators.

In a gauge theory, 22 irreducible double boxes can arise:
\begin{eqnarray}
&P^{**}_{2,2}[1],
P^{**}_{2,2}[k_{1}\cdot \ell_{2}],
P^{**}_{2,2}[(k_{1}\cdot \ell_{2})^{2}],
P^{**}_{2,2}[(k_{1}\cdot \ell_{2})^{3}],
P^{**}_{2,2}[k_{4}\cdot \ell_{1}],
\nn\\&
P^{**}_{2,2}[(k_{1}\cdot \ell_{2})(k_{4}\cdot \ell_{1})],
P^{**}_{2,2}[(k_{1}\cdot \ell_{2})^{2}(k_{4}\cdot \ell_{1})],
P^{**}_{2,2}[(k_{1}\cdot \ell_{2})^{3}(k_{4}\cdot \ell_{1})],
P^{**}_{2,2}[(k_{4}\cdot \ell_{1})^{2}],
\nn\\&
P^{**}_{2,2}[(k_{1}\cdot \ell_{2})(k_{4}\cdot \ell_{1})^{2}],
P^{**}_{2,2}[(k_{1}\cdot \ell_{2})^{2}(k_{4}\cdot \ell_{1})^{2}],
P^{**}_{2,2}[(k_{1}\cdot \ell_{2})^{3}(k_{4}\cdot \ell_{1})^{2}],
P^{**}_{2,2}[(k_{4}\cdot \ell_{1})^{3}],
\nn\\&
P^{**}_{2,2}[(k_{1}\cdot \ell_{2})(k_{4}\cdot \ell_{1})^{3}],
P^{**}_{2,2}[(k_{1}\cdot \ell_{2})^{2}(k_{4}\cdot \ell_{1})^{3}],
P^{**}_{2,2}[(k_{1}\cdot \ell_{2})^{3}(k_{4}\cdot \ell_{1})^{3}],
P^{**}_{2,2}[(k_{4}\cdot \ell_{1})^{4}],
\nn\\&
P^{**}_{2,2}[(k_{1}\cdot \ell_{2})(k_{4}\cdot \ell_{1})^{4}],
P^{**}_{2,2}[(k_{1}\cdot \ell_{2})^{2}(k_{4}\cdot \ell_{1})^{4}],
P^{**}_{2,2}[(k_{1}\cdot \ell_{2})^{4}],
P^{**}_{2,2}[(k_{1}\cdot \ell_{2})^{4}(k_{4}\cdot \ell_{1})],
\nn\\&
P^{**}_{2,2}[(k_{1}\cdot \ell_{2})^{4}(k_{4}\cdot \ell_{1})^{2}]\,,
\end{eqnarray}
where we have omitted the momentum arguments $k_{1},k_{2},k_{3},k_{4}$
for brevity.
In a gravitational theory, higher powers of the irreducible numerators
may occur.

Two of the three IBP-generating vector pairs suffice to generate all possible
IBP equations for these integrals (the third pair yields only linear
combinations of the same equations).  If we require that the coefficients
be non-vanishing in the limit $\eps\rightarrow 0$, we find 19 equations;
we can obtain an additional equation by relaxing this constraint.  This
allows us to eliminate 20 of the 22 integrals, solving for them in
terms of integrals with fewer propagators and two irreducible master
integrals, for example,
\begin{equation}
P^{**}_{2,2}[1]{\rm\ and\ }
P^{**}_{2,2}[k_1\tcdot \ell_2]\,.
\label{MasslessDoubleBoxIrreducibleMasters}
\end{equation}
This reduction is the same as previously obtained with
\AIR~\cite{AIR} (and presumably by others).
Even though the coefficients may be of order $\eps$, the solutions
do {\it not\/} involve singular coefficients for the double-box master
integrals.  (Because we do not fully reduce the simpler integrals, we
cannot determine whether that is also true for them.)  We do not
need to study the full set of 22 integrals; a minimal set that reduces
fully with the same two IBP-generating vectors is,
\begin{equation}
\{P^{**}_{2,2}[1]\,, 
P^{**}_{2,2}[k_{1}\!\cdot\!\ell_{2}]\,, 
P^{**}_{2,2}[(k_{1}\!\cdot\!\ell_{2})^{2}]\,,
 P^{**}_{2,2}[k_{4}\!\cdot\!\ell_{1}]\,, 
P^{**}_{2,2}[k_{1}\!\cdot\!\ell_{2}k_{4}\!\cdot\!\ell_{1}]\,, 
P^{**}_{2,2}[(k_{4}\!\cdot\!\ell_{1})^{2}]\}
\label{MasslessDoubleBoxMinimalSet}
\end{equation}
In addition to the IBP equation~(\ref{FirstDoubleBoxIBP}) 
or~(\ref{FirstDoubleBoxIBPsimp}), we have three additional equations
for this set,
arising from the first IBP-generating vector pair with prefactors
$k_4\tcdot\ell_1$ or $k_2\tcdot\ell_2$, and from the second IBP-generating
vector pair with no prefactor.

The form of such minimal sets will in general depend on the dimensionality
of the solution vectors; simplifying the IBP-generating vector pairs by
taking linear combinations can in general lead to simpler minimal sets
of effectively-reducible integrals.

The above reductions hold to all orders in the dimensional regulator $\eps$.
We can also ask whether additional relations appear when we drop
terms of $\Ord(\eps)$ in the integrals.  For the double box, there is
only one Gram determinant which can lead to such a relation,
\begin{equation}
\GramO{\ell_1,\ell_2,1,2,4}\,,
\label{GramDetMasslessBox}
\end{equation}
and one linear combination,
\begin{equation}
\GramO{\ell_1,1,2,4}\GramO{\ell_2,1,2,4}
-\GramSq{\ell_1,1,2,4}{\ell_2,1,2,4}\,.
\label{GramDetDifference}
\end{equation}
When we use the IBP equations to reduce
\begin{equation}
P^{**}_{2,2}\bigl[\GramO{\ell_1,\ell_2,1,2,4}\bigr]\,,
\end{equation}
however, we find that the two irreducible master 
integrals~(\ref{MasslessDoubleBoxIrreducibleMasters}) both appear with
coefficients of $\Ord(\eps)$, and hence the Gram determinant fails to
produce a useful relation.  (More precisely, it provides only a relation
for the divergent terms in the two integrals, but not for their finite
terms.)  The same is true for the combination of \eqn{GramDetDifference}.
This strongly suggests
that both integrals~(\ref{MasslessDoubleBoxIrreducibleMasters})
 that are independent to all orders in
$\e$ remain linearly independent when truncated to $\Ord(\e^0)$.
(It doesn't provide a complete proof because we have not proven
that the expressions~(\ref{GramDetMasslessBox}) 
and~(\ref{GramDetDifference}) give all possible relations of this type.)

\section{Massive Double Boxes}
\label{MassiveDoubleBoxSection}

\begin{figure}[ht]
\begin{minipage}[b]{0.4\linewidth}
\centering
\includegraphics[scale=0.7]{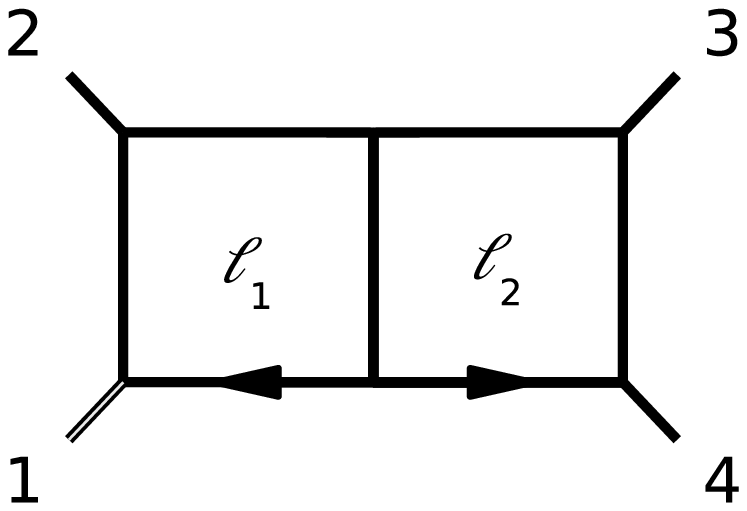}
\center{(a)}
\end{minipage}
\hspace{0.5cm}
\begin{minipage}[b]{0.4\linewidth}
\centering
\includegraphics[scale=0.7]{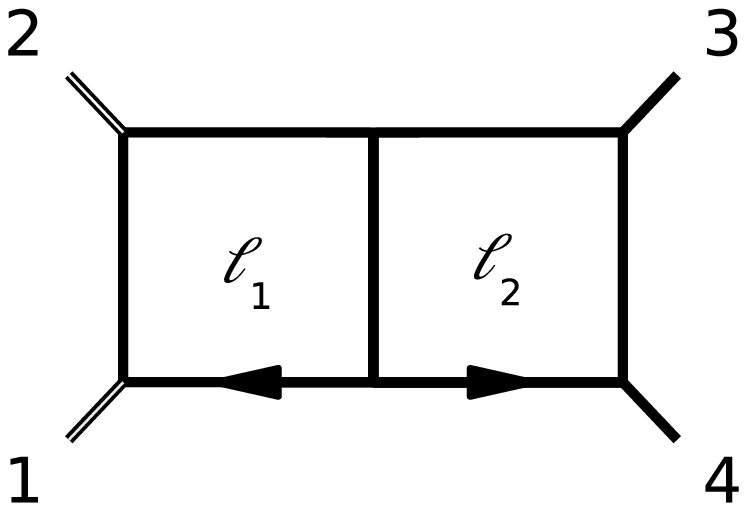}
\center{(b)}
\end{minipage}
\begin{minipage}[b]{0.4\linewidth}
\centering
\includegraphics[scale=0.7]{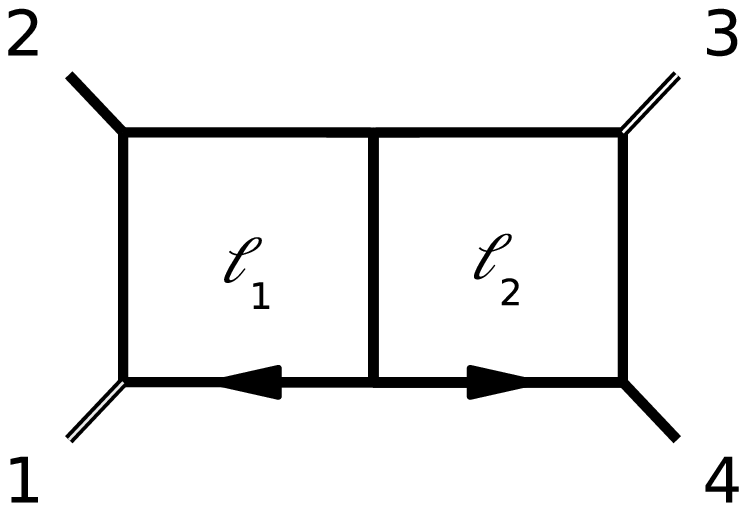}
\center{(c)}
\end{minipage}
\hspace{0.5cm}
\begin{minipage}[b]{0.4\linewidth}
\centering
\includegraphics[scale=0.7]{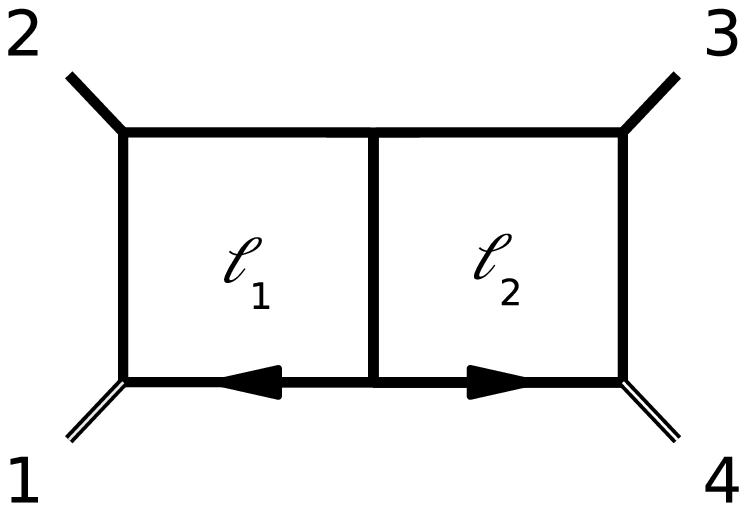}
\center{(d)}
\end{minipage}
 \caption{Double boxes with external masses, with the massive legs
indicated by doubled lines: (a) one-mass (b) short-side two-mass 
(c) diagonal two-mass (d) long-side two-mass.} 
\label{MassiveDoubleBoxesFigure}
\end{figure}

In this section, we survey the IBP-generating vectors for double boxes
with some of the external legs taken to be massive.   There is one
possible configuration of masses if one external leg is massive, as is
also true if three or four external legs are massive.  With two massive
external legs, there are three possible inequivalent integrals: both
massive legs adjacent and 
attached to the same loop (`short side' or $2{\rm m}s$);
the massive legs attached to diagonally-opposite corners 
('diagonal' or $2{\rm m}d$); or
massive legs adjacent but attached to different loops
(`long side' or $2{\rm m}l$).  The one- and two-mass double boxes
are shown in fig.~\ref{MassiveDoubleBoxesFigure}.

We can use the same basis momenta and hence same 
form~(\ref{DoubleBoxGeneralForm}) and the same basic 
symbols~(\ref{BasicVariablesDoubleBox}) as in the massless case.
Following the procedure outlined in section~\ref{IBPGeneratingSection},
we find three IBP generating vectors
for the one-mass double box (we take leg~1 to
be the massive one), once again one
with coefficients of engineering dimension two, and two with
coefficients of engineering dimension four.  
It again suffices to use the first two vectors to generate all possible 
IBP equations; there are again 20 equations for the 22 nominally-irreducible
integrals, giving rise to two irreducible master integrals, say,
\begin{equation}
P^{**}_{2,2}[1]{\rm\ and\ }
P^{**}_{2,2}[k_1\tcdot \ell_2]\,.
\label{OneMassDoubleBoxIrreducibleMasters}
\end{equation}
The set of integrals in \eqn{MasslessDoubleBoxMinimalSet}
is again a minimal set that can be reduced.
We will not display the IBP-generating vectors explicitly, but they
are given in a companion {\sl Mathematica\/} file.

For the long-side two-mass double box, we take legs~1 and~4 to be massive,
and now find five IBP generating vectors, all with coefficients
of engineering dimension four.  There are again 20 equations for
the 22 nominally-irreducible integrals, which we can
derive using three of the five pairs of vectors.
We can again pick 
the integrals in \eqn{OneMassDoubleBoxIrreducibleMasters} as irreducible
masters.

For the diagonal two-mass double box, we take legs~1 and~3 to be massive,
and use algorithm~III to
find three IBP generating vectors, with the same dimensions
as the massless and one-mass cases.  Once again, we need
to use only two vector pairs to generate all required equations, 
and can take the
integrals in~\eqn{OneMassDoubleBoxIrreducibleMasters} as irreducible
masters.

When we examine the short-side two-mass double box (taking
legs~1 and~2 to be massive), we find our first 
surprise.  Here we find four IBP generating vectors,
all of engineering dimension four; but we find only 19 equations
for the 22 original integrals (for which we need three of
the four vector pairs), leaving us with three irreducible
master integrals,
\begin{equation}
P^{**}_{2,2}[1]\,,
P^{**}_{2,2}[k_1\tcdot \ell_2]\,,
{\rm\ and\ }
P^{**}_{2,2}[k_4\tcdot \ell_1]\,.
\label{TwoMassDoubleBoxIrreducibleMasters}
\end{equation}

The three- and four-mass double boxes lead to very complicated
analytic expressions in intermediate stages; it is much faster (and
sufficient for the unitarity approach, as discussed earlier), 
to compute
the IBP-generating vectors for a fixed numerical configuration of
external momenta.  We have chosen to do so; 
we find five IBP-generating
vectors for the three-mass case (three with coefficients of dimension four,
and two with coefficients of dimension six), and four vector pairs
for the four-mass case (two each of dimensions four and six).
In line with the result
for the short-side two-mass double box, 
we find three master integrals for the three-mass case
(which we can take to be those in \eqn{TwoMassDoubleBoxIrreducibleMasters}).
For the four-mass case, we find that we need four master integrals,
which we can take to be,
\begin{equation}
P^{**}_{2,2}[1]\,,
P^{**}_{2,2}[k_1\tcdot \ell_2]\,,
P^{**}_{2,2}[k_4\tcdot \ell_1]\,,
{\rm\ and\ }
P^{**}_{2,2}[k_1\tcdot \ell_2 k_4\tcdot \ell_1]\,.
\label{FourMassDoubleBoxIrreducibleMasters}
\end{equation}

In all cases, there are no additional equations that arise from
truncation to $\Ord(\e)$. 

\section{The Pentabox}
\label{PentaboxSection}

\begin{figure}[ht]
\includegraphics[scale=0.7]{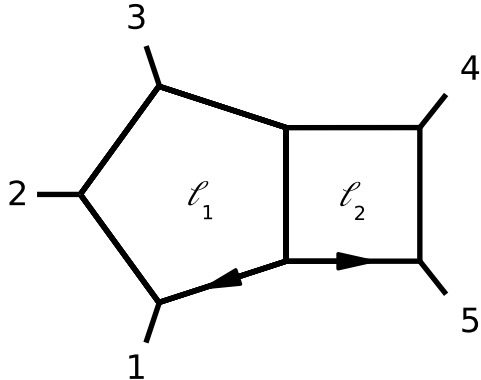}
 \caption{The pentabox $P^{**}_{3,2}$.} 
\label{PentaBoxFigure}
\end{figure}
Our next example is one of the three basic topologies that arise in five-point
computations: the pentabox $P_{3,2}^{**}$, shown in fig.~\ref{PentaBoxFigure}.
  Here, we choose
$k_{1,2,3,5}$ as basis momenta, so the general 
form~(\ref{IBPVectorGeneralForm}) becomes,
\begin{equation}
v_i^\mu = \coeffA{\ell_1}_i \ell_1^\mu
+\coeffA{\ell_2}_i \ell_2^\mu
+\coeffA{1}_i k_1^\mu
+\coeffA{2}_i k_2^\mu
+\coeffA{3}_i k_3^\mu
+\coeffA{5}_i k_5^\mu\,,
\end{equation}
where again each of the coefficients $\coeffA{p}$ is a function
of Lorentz invariants in the set of symbols $V_{32}$,
\begin{equation}
V_{32} = \{
\ell_1^2\,, \ell_1\cdot \ell_2, \ell_2^2\,,
\ell_1\cdot k_1\,,
\ell_1\cdot k_2\,,
\ell_1\cdot k_3\,,
\ell_1\cdot k_5\,,
\ell_2\cdot k_1\,,
\ell_2\cdot k_2\,,
\ell_2\cdot k_4\,,
\ell_2\cdot k_5\,,s_{12}\}\,.
\label{BasicVariablesPentabox}
\end{equation}

For this integral, we have constructed vectors both analytically and
numerically; the numerical construction is much less memory-consuming.
In both cases, the algorithms yield six IBP-generating vectors with
coefficients of engineering dimension four, and three vectors with
coefficients of dimension six.  Their forms are too lengthy
to display in the text, but are provided in the companion 
{\it Mathematica\/} file.  There are 76
nominally-irreducible integrals in a gauge theory, involving 
powers of the three irreducible numerators,
\begin{equation}
k_1\tcdot\ell_2\,, k_2\tcdot\ell_2\,, k_5\tcdot\ell_1\,.
\end{equation}
It suffices to use
the six vector pairs of dimension four to generate all possible equations
for these integrals.  We find 73 such equations, leaving us with three
truly-irreducible master integrals, which we can choose to be,
\begin{equation}
P^{**}_{3,2}[1]\,,
P^{**}_{3,2}[k_1\tcdot \ell_2]\,,
P^{**}_{3,2}[k_5\tcdot \ell_1]\,.
\label{PentaboxMasterIntegrals}
\end{equation}

Examples of these reduction equations are,
\begin{eqnarray}
P^{**}_{3,2}[k_{2}\!\cdot\!\ell_{2}]
&=&
-\frac{ (\chi_{15}-2 \chi_{23}+\chi_{23}\chi_{34}+2 \chi_{45}+\chi_{15}\chi_{45}- \chi_{34}\chi_{45})s_{12}}{4 (\chi_{15}- \chi_{23}+\chi_{45})}
   P^{**}_{3,2}[1]\nn\\
&& 
-\frac{(1+\chi_{15}- \chi_{23}- \chi_{34})}{\chi_{15}- \chi_{23}+\chi_{45}}
   P^{**}_{3,2}[k_{1}\!\cdot\!\ell_{2}]
  +{\rm\ simpler\ integrals}\,,\nn\\
P^{**}_{3,2}[k_{1}\!\cdot\!\ell_{2}k_{2}\!\cdot\!\ell_{2}]
&=&
 \chi_{15}\biggl(\frac{1+\chi_{15}-\chi_{34}-\chi_{45}-\chi_{15}\chi_{45}+\chi_{23}\chi_{45}+\chi_{34}\chi_{45}}{8 (1-2 \e)(1-\chi_{34}-\chi_{45})}\nn\\
&&\hskip 10mm+\frac{\e(1+\chi_{15}-\chi_{23}-\chi_{34})}
   {8 (1-2 \e)(\chi_{15}-\chi_{23}+\chi_{45})(1-\chi_{34}-\chi_{45})}\nn\\
&&\hskip 15mm\times
\bigl(\chi_{15}(1-\chi_{45})+(\chi_{45}-\chi_{23})(2-\chi_{34}-2 \chi_{45})
\bigr)
\biggr)
s_{12}^{2}P^{**}_{3,2}[1]\nn\\
&&+\biggl(\frac{\e(\chi_{15}+2 \chi_{15}\chi_{23}-2 \chi_{23}^{2}-\chi_{23}\chi_{34}-\chi_{15}\chi_{45}+2 \chi_{23}\chi_{45}+\chi_{34}\chi_{45})}{2 (1-2 \e)(\chi_{15}-\chi_{23}+\chi_{45})}\nn\\
&&-\frac{2+2 \chi_{15}-3 \chi_{34}-\chi_{15}\chi_{34}+\chi_{34}^{2}-2 \chi_{45}-2 \chi_{15}\chi_{45}+\chi_{23}\chi_{45}+2 \chi_{34}\chi_{45}}{2 (1-2 \e)(1-\chi_{34}-\chi_{45})}\biggr)\nn\\
&&\hskip 7mm\times s_{12}P^{**}_{3,2}[k_{1}\!\cdot\!\ell_{2}]\nn\\
&&-\frac{(1+2\e)(1+\chi_{15}-\chi_{23}-\chi_{34})(1+\chi_{23}-\chi_{45})}
   {4 (1-2 \e)(1-\chi_{34}-\chi_{45})}
   s_{12}P^{**}_{3,2}[k_{5}\!\cdot\!\ell_{1}]\nn\\
&&+{\rm\ simpler\ integrals}\,.
\end{eqnarray}

These master integrals are independent when considered
to all orders in $\e$.  Unlike
the case of the double box, however, here we {\it can\/} find two linear
relations between them, so long as we truncate at $\Ord(\e^0)$.
These two relations arise from considering the following two integrals,
\begin{equation}
P^{**}_{3,2}\biggl[\Gram{\ell_1,1,2,3,5}{\ell_2,1,2,3,5}\biggr]
{\rm\ and\ }
P^{**}_{3,2}\biggl[k_5\tcdot\ell_1 \Gram{\ell_1,1,2,3,5}{\ell_2,1,2,3,5}\biggr]
\end{equation}
both of which are of $\Ord(\e)$, as discussed in 
\sect{AdditionalIdentitiesSubsection}. Setting the two to zero,
and using the reductions obtained from integration by parts, we find two
equations relating the masters in~\eqn{PentaboxMasterIntegrals}.  We
can use these, for example, to eliminate the two integrals with non-trivial
numerators in favor of $P^{**}_{3,2}[1]$,
\begin{eqnarray}
P^{**}_{3,2}[k_{5}\!\cdot\!\ell_{1}] 
&=&
\frac{\chi_{15}\chi_{34}\chi_{45}s_{12}}{-\chi_{15}+\chi_{23}-\chi_{23}\chi_{34}+\chi_{15}\chi_{45}+\chi_{34}\chi_{45}}P^{**}_{3,2}[1]
+{\rm\ simpler\ integrals}+\Ord(\e)\,,\nn\\
P^{**}_{3,2}[k_{1}\!\cdot\!\ell_{2}]
&=&
 \frac{
 \chi_{15}(\chi_{15}(1-\chi_{45})^{2}+\chi_{34}(1-\chi_{45})\chi_{45}
  -\chi_{23}(1-\chi_{45}-\chi_{34}(1+\chi_{45})))
  }
 {4 (1-\chi_{34}-\chi_{45})(\chi_{15}-\chi_{23}+\chi_{23}\chi_{34}-\chi_{15}\chi_{45}-\chi_{34}\chi_{45})} 
   s_{12}P^{**}_{3,2}[1]\nn\\
&&+{\rm\ simpler\ integrals}+\Ord(\e)\,.
\end{eqnarray}
  The other Gram determinants
of the form suggested in \sect{AdditionalIdentitiesSubsection}
do not yield independent equations, but could be used instead to
obtain equivalent equations.  The longer denominators in these expressions
may develop poles at exceptional values of the kinematics; these are
presumably spurious and are softened by the behavior of the
various integrals in those limits, but we have not checked this.  (Other
denominators vanish in non-adjacent collinear limits, for example 
$1-\chi_{34}-\chi_{45} \rightarrow 0$ in the collinear limit 
$k_3\parallel k_5$; these are presumably spurious as well.)

One may wonder whether the IBP equations are even required for reduction
of the truncated integrals, given the seemingly-stronger equations
arising from Gram determinants.  However, this strength is illusory: if
we use only Gram determinant equations (including the IBP-like ones
built from Gram determinants of the form given in \eqn{EpsGramDetsIII}),
we find only 24 equations for 29 of the 35 integrals with numerators of
dimension eight or less.  (The IBP-like determinants provide two of
these equations.)  This would leave five seemingly-irreducible
integrals as masters; of course, using the IBP equations, we could
then reduce all of these to the scalar integral $P^{**}_{3,2}[1]$.
If we consider the complete set of 76 integrals, we see another
problem with using Gram-determinant equations alone: 20 of the integrals
(those with four powers of $\ell_2$) are ultraviolet divergent, which
prevents us from using these equations to simplify them.  In addition,
even amongst the ultraviolet-finite integrals, we are left with
five master integrals (there are 51 equations in total that we could
derive).

\section{A Six-Point Example}
\label{SixPointSection}

\begin{figure}[ht]
\includegraphics[scale=0.7]{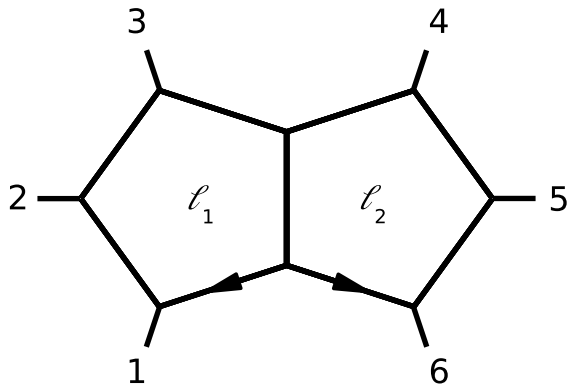}
 \caption{The double pentagon $P^{**}_{3,3}$.} 
\label{DoublePentagonFigure}
\end{figure}
We will consider one example of a six-point integral, the so-called
double pentagon $P^{**}_{3,3}$, shown in fig.~\ref{DoublePentagonFigure}.
In contrast to the pentabox $P^{**}_{3,2}$ considered in the 
previous section, we find that this integral can be reduced
to simpler integrals
entirely using Gram determinant equations alone.  Indeed, not only
can integrals with non-trivial numerators be reduced, but the 
scalar integral itself, $P^{**}_{3,3}[1]$, can also be expressed
in terms of simpler integrals (pentaboxes and products of one-loop
pentagons, themselves reducible) via
algebraic identities.  We don't even need those Gram determinants equivalent
to IBP equations to perform these reductions.

The double pentagon has two irreducible numerators, which 
we can pick to be $k_6\cdot\ell_1$
and $k_1\cdot\ell_2$.  There are thus 33 formally-irreducible integrals
that arise in a gauge theory.  The first thing to notice is that they
are all ultraviolet-finite, so one of the obstructions that existed
in the pentabox case to use of Gram-determinant equations alone for
a complete reduction is absent here.  
There are 15 integrals of engineering dimension eight or less;
exclude $P^{**}_{3,3}[(k_6\tcdot\ell_1)^4]$ and 
$P^{**}_{3,3}[(k_1\tcdot\ell_2)^4]$, and examine the remaining 13
integrals.  We can find 13 independent equations for them by 
starting with the following identities,
\begin{equation}
\Ord(\e) = P^{**}_{3,3}\biggl[p\,\Gram{\ell_1,1,2,3,6}{\ell_2,1,2,3,6}\biggr]\,,
\end{equation}
with prefactors $p=1, k_1\tcdot\ell_2, (k_1\tcdot\ell_2)^2,
k_6\tcdot\ell_1,(k_6\tcdot\ell_1)^2,
k_1\tcdot\ell_2\,k_6\tcdot\ell_1$;
\begin{equation}
\Ord(\e) = P^{**}_{3,3}\biggl[p\,\Gram{\ell_1,\ell_2,1,2,3}{\ell_1,\ell_2,4,5,6}\biggr]\,,
\end{equation}
with prefactors $p=1, k_1\tcdot\ell_2, (k_1\tcdot\ell_2)^2,
k_6\tcdot\ell_1, (k_6\tcdot\ell_1)^2$;
and 
\begin{equation}
\Ord(\e) = P^{**}_{3,3}\biggl[\Gram{\ell_1,\ell_2,1,2,4}{\ell_1,\ell_2,3,5,6}\biggr]
\quad {\rm\ and\ }\quad
\Ord(\e) = P^{**}_{3,3}\biggl[\Gram{\ell_1,\ell_2,1,2,5}{\ell_1,\ell_2,3,4,6}\biggr]\,.
\end{equation}

Similarly, if we examine the full set of 33 formally-irreducible
integrals, we find 33 independent equations.

\section{Connection to Generalized Unitarity}
\label{GeneralizedUnitaritySection}

In this section, we use generalized unitarity to give a heuristic
explanation for the structure of the results presented in previous
sections.  We begin, as in section~\ref{OneLoopSection}, with a
discussion at one loop.

In basic unitarity at one loop, we examine the branch cut of
amplitudes channel by channel.  In each channel, the branch is a
phase-space integral over a product of tree amplitudes.  In the
present paper, we are focused on loop integrals rather than complete
amplitudes, so the equivalent statement --- dating back to the
Cutkosky rules~\cite{Cutkosky}
 of the 1960s --- is the expression of the branch cut in
terms of phase space integrals of scalar tree diagrams.  The ordinary
cut may be obtained by cutting two propagators, that is replacing the
propagators by positive-energy delta functions which put the
intermediate state on shell,
\begin{equation}
{i\over (\ell-K)^2+i\varepsilon} \longrightarrow 
2 \pi \delta^{(+)}\bigl( (\ell-K)^2\bigr)\,.
\end{equation}
(The $(+)$ superscript indicates the restriction to positive energies.)
There is nothing stopping us, however, from cutting more than two
propagators, and this is the idea behind generalized unitarity.  The
solutions to the delta function constraints will in general then be
complex, and so the delta functions must be understood in a more
general sense, as contour integrals with the contours chosen to encircle
the common solutions to the constraint equations.  
The idea of generalized unitarity was first applied as
a practical tool for computation of amplitudes by Bern, Dixon, and one
of the authors~\cite{Zqqgg}.  It was later combined with the use of
complex momenta to give a general algebraic solution to finding the
coefficients of box integrals~\cite{BCFUnitarity}, and used to derive a general
and numerically-applicable technique for triangle and bubble integrals
by Forde~\cite{Forde}.

If one cuts as many propagators as possible, one arrives at maximal
unitarity, as used for example in ref.~\cite{FiveLoop,MaximalUnitarity}.  In
old-fashioned language, this is equivalent to looking for `leading
singularities', discussed in a modern incarnation in
ref.~\cite{LeadingSingularity}.

But how many propagators {\it can\/} we cut?  If we examine a one-loop
amplitude with all external momenta taken to be massive (so that
infrared singularities are tamed), we can take the dimensional
regulator $\e$ to zero, and perform the integrals in four dimensions.
(Ignore the ultraviolet-divergent bubble in this discussion.)  Each
delta function will impose one constraint; because we have four
components, we can have up to four delta functions.  Attempting to
impose additional delta functions will in general yield no solutions.
(More precisely, because we will have more delta functions than
integrals, the result will itself be a delta function rather than an
ordinary function.)  This in turn implies that functions with
additional propagators may be determined in terms of functions with up
to four propagators, as there are no additional degrees of freedom.
In this case, all pentagons or higher-point integrals are reducible to
sums of boxes and lower-point integrals.

The generalization of this observation to higher loops is
straightforward.  At each loop order, we have an additional four
components.  We can thus cut an additional four propagators.  When
considering infrared- and ultraviolet-finite integrals, then, we
expect that only those with up to four propagators per loop momentum
will be algebraically independent.  At two loops, this means that
integrals with more than eight propagators, or more than four
propagators involving a single loop momentum, will be reducible into
simpler integrals.  

Of course, the integrals of interest are in general infrared
divergent.  While the loop momentum is formally $D$-dimensional, so
long as we keep the external momenta in four dimensions, the
additional components $\mu$ can only appear at one loop as $\mu^2$, on
which we can impose one additional delta function.  Thus when
considering one-loop integrals to all orders in $\e$, the pentagon
integral must be taken as an additional independent integral, while
higher-point integrals remain reducible.  Now, the algebraic
independence of the pentagon only manifests itself at $\Ord(\e)$;
terms through $\Ord(\e^0)$ {\it are\/} reducible to sums of boxes.
This reducibility is not manifest in our heuristic discussion; but it
suggests the conjecture that the reducibility of integrals to
$\Ord(\e^0)$ follows the pattern of massive reductions.

What happens at two loops?  We now have two loop momenta, and
correspondingly two $\e$-dimensional vectors, $\mu_1$ and $\mu_2$.
These can now appear in integrals in the form of three independent
quantities, $\mu_1^2$, $\mu_2^2$, and $\mu_1\cdot \mu_2$.  We could
impose additional delta functions on each, corresponding to cutting
three additional propagators.  For two-loop diagrams, we therefore
expect any integral with more than eleven propagators, or more than
five propagators involving a lone loop momentum, to be reducible to
all orders in $\e$.  This is exactly what we found in
section~\ref{PlanarSection}.

\def\dash{\hbox{-\kern-.02em}}
Different propagators lead to different branch points (or
branch surfaces, for the many-complex-variables functions
we are considering).  Accordingly, the algebraic
independence of uncut propagators is clear.
The algebraic independence of non-trivial numerators is 
less clear, as one might imagine algebraic relations
between them.  (Indeed, there are clearly algebraic
relations between different powers of numerators, as
seen in reduction equations elsewhere in the literature
or in previous sections.)  Heuristically, we do at least
expect an upper bound, 
\begin{equation}
\# {\rm\ irreducible\ non\dash{}trivial\ numerator\ integrals}
\leq 11 - \#{\rm propagators}\,.
\end{equation}
 This bound is respected by the
explicit results in previous sections.  

The question of algebraic independence when truncating integrals to
$\Ord(\e)$ is more subtle.  If we adopt the conjecture above suggested
by one-loop results, it would imply that only truncated integrals with
up to eight propagators (and up to four involving each loop momentum)
are algebraically independent.  This is in agreement with the
reducibility of the double-pentagon $P^{**}_{3,3}$ discussed in
section~\ref{SixPointSection}.   We might be further tempted to conjecture
that the number of independent integrals with non-trivial numerators
is limited to eight less the number of propagators.  This would
imply that there are no independent pentaboxes with irreducible
numerators, which is in fact true.
Thus the bound is
respected by the pentabox results discussed in section~\ref{PentaboxSection},
and also by the results for some double boxes; but it is violated for
the short-side two-mass double box, as well as for double boxes with three
or four external masses.  The precise manner in which the heuristic picture
breaks down remains to be clarified.

\section{Conclusions}
\label{ConclusionsSection}

Knowledge of an integral basis plays an important role in modern
unitarity calculations.  In this paper, we have given an outline
of a basis for planar integrals (with massless propagators)
at two loops.  We distinguish two
kinds of bases: the first, 
a set of integrals which are linearly
independent to all orders in the dimensional regulator $\eps$; the
second, in which linear independence is required only through $\Ord(\eps^0)$.

Smirnov and Petukhov~\cite{SmirnovPetukhov}
have recently shown that the integral basis
resulting from integration by parts is finite.
We have delineated an explicit finite set of integrals which contains a
minimal basis, and given an explicit procedure reducing an arbitrary
planar two-loop gauge-theory integral to an element of this set.
The set contains
only integrals with four or fewer external legs attached to each line
in the vacuum graph.  All irreducible numerators, whose number depends
on the external legs, are allowed in this finite set of integrals.  The
final basis contain only a subset of these integrals.

In order to reduce the set further,
we then introduced an approach to generating integration-by-parts equations
which involve only integrals in the desired set, along with simpler integrals
(with propagators omitted), and avoiding integrals which are not
ordinary Feynman integrals.  For each of the integrals in the above set,
one can solve for these vectors, and then determine the set of independent
master integrals that make up the first, $D$-dimensional, basis.  
Unlike the situation at one loop,
the reductions, and more importantly, the number of independent integrals,
depend on the masses of the external legs.
We gave
a few examples of this type of calculation, but leave a complete study of
the integrals to future work.

We also introduced a special set of numerators, built using Gram determinants,
which provide equations that yield identities for integrals truncated
to $\Ord(\eps^0)$.  These equations reduce the $D$-dimensional set of
independent master integrals to a smaller set making up the second,
``regulated four-dimensional'' basis.

The general arguments as well as the notion of IBP-generating vectors
and additional $\Ord(\eps)$ identities generalize to non-planar
integrals as well as to higher-loop integrals.  We also expect them to
generalize from the massless propagators considered here to integrals
with massive propagators.  It would also be interesting to explore the
analog of the $\Ord(\eps)$ identities for integrals in two- and
three-dimensional field theories.  We gave a heuristic argument for
understanding the basis in terms of generalized unitarity; it would be
interesting if it could be developed further to a complementary
derivation for the reduction of integrals with formally-irreducible
numerators to an independent set of master integrals.

The defining equations~(\ref{IBPMatrixEquation}) for the
IBP-generating vectors can also be thought of as defining a `surface'
or variety in the space whose coordinates are given by the different
monomials in $V$~(\ref{BasicVariables}).  It would be interesting to
explore its connection with the Grassmannians~\cite{Grassmannians}
introduced in recent explorations of integral coefficients in the
$\NeqFour$ theory.  The integral basis appropriate for the $\NeqFour$
theory should presumably make manifest (up to infrared divergences)
its extended symmetries (conformal and dual conformal
symmetries~\cite{DualConformalI,DualConformalII}), and may make
natural use of twistorial integrands such as those discussed in
refs.~\cite{TwistorialIntegrands}.

Our approach to solving the required equations~(\ref{IBPBasicVectorEquation})
 for the IBP-generating made use of Gr\"obner bases, and in particular
the standard Buchberger algorithm~\cite{Buchberger,AdamsLoustaunau}
 for computing them.  The present implementation of the algorithm
(coded in {\sl Mathematica\/})
performs well for simple cases like the double box, but slows down and
requires large memory in its intermediate stages for integrals with
more legs or many massive legs.  It would be worthwhile exploring the
use of more modern algorithms, such as those of Faug\`ere~\cite{Faugere}
for computing the required Gr\"obner bases.

\section*{Acknowledgments}

DAK would like to thank Zvi Bern, Thomas Gehrmann, Nigel Glover,
and Radu Roiban
 for helpful discussions.
JG would like to thank Micha\l{} Czakon for the opportunity to use his 
unpublished IdSolver package for some internal tests, and Tord Riemann 
and Bas Tausk for helpful discussions.
We also
thank Academic Technology Services at UCLA for computer support.  
This early stages of this work were supported in part by the EGIDE
program of the French Foreign Ministry, 
under grant 12516NC.  JG's
research is supported by 
the Polish Ministry of Science and High Education
from its science budget for 2010--2013 under grant number N~N202~102638, 
and also by the European research training networks 
MRTN--CT--2006--035505 ``HEPTOOLS'' and
MRTN--CT--2006--035482 ``FLAVIAnet''.
DAK's research is supported by the European
Research Council under Advanced Investigator Grant ERC--AdG--228301.


\end{document}